\documentclass[10pt,twocolumn,twoside]{IEEEtran}
\usepackage{cite}

\ifCLASSINFOpdf
\else
  \usepackage[dvips]{graphicx}
  \graphicspath{{figures/}}
  \DeclareGraphicsExtensions{.eps}
\fi
\usepackage[cmex10,tbtags]{amsmath}
\usepackage{amssymb}

\usepackage{algorithmic}
\usepackage[caption=false,font=footnotesize]{subfig}
\usepackage{float}

\usepackage{fixltx2e}
\usepackage{url}

\usepackage{psfrag}

\newcommand{\lbr}{\left[}
\newcommand{\rbr}{\right]}

\newcommand{\defeq}{\stackrel{\Delta}{=}}

\DeclareMathOperator{\sinc}{sinc}

\DeclareMathOperator*{\argmax}{arg\,max}

\DeclareMathOperator{\var}{var}

\floatstyle{ruled}
\newfloat{algorithmDSW}{tbp}{loa}
\floatname{algorithmDSW}{Algorithm}

\begin{document}
%
\title{Bayesian Post-Processing Methods for Jitter Mitigation in Sampling}
%
%
%

\author{Daniel~S.~Weller*,~\IEEEmembership{Student Member,~IEEE,}
        and~Vivek~K~Goyal,~\IEEEmembership{Senior Member,~IEEE}
\thanks{This work was supported in part by the Office of Naval Research through a National Defense Science and Engineering Graduate (NDSEG) fellowship, NSF CAREER Grant CCF-0643836, and Analog Devices, Inc.}
\thanks{D. S. Weller is with the Massachusetts Institute of Technology, Room 36-680, 77 Massachusetts Avenue, Cambridge, MA 02139 USA (phone: +1.617.324.5862; fax: +1.617.324.4290; email: dweller@mit.edu), and V. K. Goyal is with the Massachusetts Institute of Technology, Room 36-690, 77 Massachusetts Avenue, Cambridge, MA, 02139 USA (phone: +1.617.324.0367; fax: +1.617.324.4290; e-mail: vgoyal@mit.edu).}}

%
%

\markboth{Draft}{Bayesian Post-Processing Methods for Jitter Mitigation in Sampling}
%



\maketitle

\begin{abstract}
Minimum mean squared error (MMSE) estimators of signals from samples corrupted by jitter (timing noise) and additive noise are nonlinear, even when the signal prior and additive noise have normal distributions. This paper develops a stochastic algorithm based on Gibbs sampling and slice sampling to approximate the optimal MMSE estimator in this Bayesian formulation. Simulations demonstrate that this nonlinear algorithm can improve significantly upon the linear MMSE estimator, as well as the EM algorithm approximation to the maximum likelihood (ML) estimator used in classical estimation. Effective off-chip post-processing to mitigate jitter enables greater jitter to be tolerated, potentially reducing on-chip ADC power consumption.
\end{abstract}


%
\IEEEpeerreviewmaketitle

\section{Introduction}\label{sec:intro}

Reducing the power consumption of analog-to-digital converters (ADCs) would improve the capabilities of power-constrained devices like medical implants, wireless sensors, and cellular phones. Clock circuits that produce jittered (noisy) sample times naturally consume less power than those with low phase noise, so allowing high phase noise is one avenue to reduce power consumption. However, increasing jitter in an ADC reduces the effective number of bits (ENOB) (rms accuracy on a dyadic scale) by one for every doubling of the jitter standard deviation, as described in~\cite{Walden99} and~\cite{Brannon00}. Compensating for the reduced ENOB by designing more accurate comparators increases power consumption by a factor of four for every lost bit of accuracy~\cite{Uyttenhove01}. Thus, to achieve reduced on-chip power consumption, the lost bits should be recovered in a different manner.

In~\cite{WellerClassical}, the authors post-process the jittered samples, employing an EM algorithm to perform classical maximum likelihood (ML) estimation of the signal parameters. This nonlinear classical estimation technique is capable of tolerating between $1.4$ and $2$ times the jitter standard deviation that can be mitigated by linear estimation. In this work, nonlinear post-processing is extended to the Bayesian framework, where the signal parameters are estimated knowing their prior distribution. Here, we do not require that signal and noise variances are known a priori; our hierarchical Bayesian model includes prior distributions on these parameters. The technique presented here achieves significant improvement over linear estimation for a wider range of jitter variance than the EM algorithm from~\cite{WellerClassical}, improving the applicability of nonlinear post-processing.

The block post-processing of the jittered samples is intended to be performed off-chip (e.g.\ on a PC), so we do not attempt to optimize the total power consumption, including the digital post-processing.  However, we are concerned with making prudent choices in algorithm design so that the computational complexity of the post-processing is reasonable.  The problem of mitigating jitter also can be motivated by loosening manufacturing tolerances (hence reducing cost) or by problems in which spatial locations of sensors are analogous to sampling times~\cite{Deshpande10}.

\subsection{Problem Formulation}

Consider the shift-invariant subspace of $L^2(\mathbb{R})$ associated with a generating function $h(t)$ and a signal $x(t)$ in that subspace:
\begin{equation}
x(t) = \sum_{k\in\mathbb{Z}} x_kh(t/T-k).\label{eq:intro_sisubspace}
\end{equation}
Assuming $\{h(t/T-k) : k\in\mathbb{Z}\}$ is a Riesz basis for the subspace, $x(t)$ is in one-to-one correspondence with the sequence $\{x_k\}_{k \in \mathbb{Z}}$ and the sequence $\{x_k\}_{k \in \mathbb{Z}}$ is in $\ell^2(\mathbb{Z})$. Examples of $h(t)$ include the function $\sinc(t) \defeq \frac{\sin(\pi t)}{\pi t}$ used throughout this paper, as well as B-splines and wavelet scaling functions as discussed in~\cite{Unser00}. While $h(t) = \sinc(t)$ is used for simulations, the developments in this paper are not specialized to the form of $h(t)$ in any way. We only require that $h(t)$ satisfies the Riesz basis condition and that the sampling prefilter $s(-t)$ satisfies the biorthogonality condition $\langle h(t/T-k),\, s(t/T-\ell) \rangle = \delta_{k-\ell}$, for all $k,\,\ell \in \mathbb{Z}$. The Riesz basis condition allows bounding of $L^2$ error of $x(t)$ in terms of $\ell^2$ error of $x_k$; when $\{h(t/T-k) : k\in\mathbb{Z}\}$ is an orthogonal set, these errors are constant multiples. When $h(t) = \sinc(t)$, the shift-invariant subspace is the subspace of signals with Nyquist sampling period $T$. Without loss of generality, we assume $T = 1$.

\begin{figure}[!t]
\centering
\psfrag{x(t)}[][]{$x(t)$}
\psfrag{tn}[][]{$t_n$}
\psfrag{s(t)}[][]{$s(-t)$}
\psfrag{wn}[][]{$w_n$}
\psfrag{ADC}[][]{ADC}
\psfrag{y}[][]{$\mathbf{y}$}
\psfrag{est}[][]{estimator}
\psfrag{px}[][]{$p(\mathbf{x},\sigma_x^2)$, $p(\sigma_z^2)$, $p(\sigma_w^2)$}
\psfrag{offchip}[][]{off-chip}
\psfrag{out}[][]{$\mathbf{\hat{x}}$}
\includegraphics[width=3.45in]{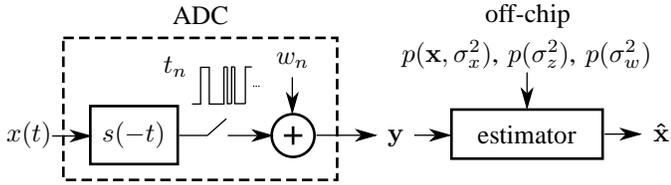}
\caption{Block diagram of an abstract ADC with off-chip post-processing. The signal $x(t)$ is filtered by the sampling prefilter $s(-t)$ and sampled at time $t_n$. These samples are corrupted by additive noise $w_n$ to yield $y_n$. The post-processor estimates the parameters $\mathbf{x}$ of $x(t)$ using the vector of $N$ samples $\mathbf{y}$ from the ADC.}
\label{fig:adcblockdgrm}
\end{figure}
When observing the signal $x(t)$ through a sampling system like an ADC, the analog signal is prefiltered by $s(-t)$, and samples $y_n$ are taken of the result at jittered times $t_n = nT_s + z_n$. To model oversampled ADCs, we oversample the signal by a factor of $M$, so the sampling period is $T_s = 1/M$. The samples are also corrupted by additive noise $w_n$, which models auxiliary effects like quantization and thermal noise. For $h(t) = \sinc(t)$, the dual $s(t) = \sinc(t)$ is an ideal lowpass filter with bandwidth $2\pi$. The observation model, depicted in Figure~\ref{fig:adcblockdgrm}, is
\begin{equation}
y_n = \left[x(t) * s(-t)\right]_{t = \frac{n}{M}+z_n} + w_n.
\end{equation}

We aim to estimate a block of $K$ coefficients, assuming the remaining coefficients are negligible:
\begin{equation}
x(t) \approx \sum_{k=0}^{K-1} x_kh(t-k).\label{eq:intro_sigmodel}
\end{equation}
This specializes the observation model to
\begin{equation}
y_n = \sum_{k=0}^{K-1} x_kh\left(\frac{n}{M}+z_n-k\right) + w_n.\label{eq:intro_obsmodel}
\end{equation}
Grouping the variables into vectors, let $\mathbf{x} = [x_0,\ldots,x_{K-1}]^T$, $\mathbf{y} = [y_0,\ldots,y_{N-1}]^T$, $\mathbf{z} = [z_0,\ldots,z_{N-1}]^T$, and $\mathbf{w} = [w_0,\ldots,w_{N-1}]^T$. Then, in matrix form,
\begin{equation}
\mathbf{y} = \mathbf{H(z)x+w},\label{eq:intro_obsmodel2}
\end{equation}
where $[\mathbf{H(z)}]_{n,k} = h(\frac{n}{M}+z_n-k)$, for $n = \{0,\ldots,N-1\}$, and $k = \{0,\ldots,K-1\}$. Let $\mathbf{h}_n^T(z_n)$ be the $n$th row of $\mathbf{H(z)}$. Also, denote the $k$th column of $\mathbf{H(z)}$ by $\mathbf{H}_k(\mathbf{z})$ and the matrix with the remaining $K-1$ columns by $\mathbf{H}_{\backslash k}(\mathbf{z})$. Similarly, let $\mathbf{x}_{\backslash k} = [x_0,\ldots,x_{k-1},x_{k+1},\ldots,x_{K-1}]^T$ be the vector of all but the $k$th signal coefficient.

\begin{figure}[!t]
\centering
\psfrag{ax}[][]{$\alpha_x$}
\psfrag{bx}[][]{$\beta_x$}
\psfrag{sx2}[][]{$\sigma_x^2$}
\psfrag{az}[][]{$\alpha_z$}
\psfrag{bz}[][]{$\beta_z$}
\psfrag{sz2}[][]{$\sigma_z^2$}
\psfrag{aw}[][]{$\alpha_w$}
\psfrag{bw}[][]{$\beta_w$}
\psfrag{sw2}[][]{$\sigma_w^2$}
\psfrag{x0}[][]{$x_0$}
\psfrag{xk}[][]{$x_k$}
\psfrag{xK-1}[][]{$x_{K-1}$}
\psfrag{zn}[][]{$z_n$}
\psfrag{wn}[][]{$w_n$}
\psfrag{yn}[][]{$y_n$}
\includegraphics[width=3.2in]{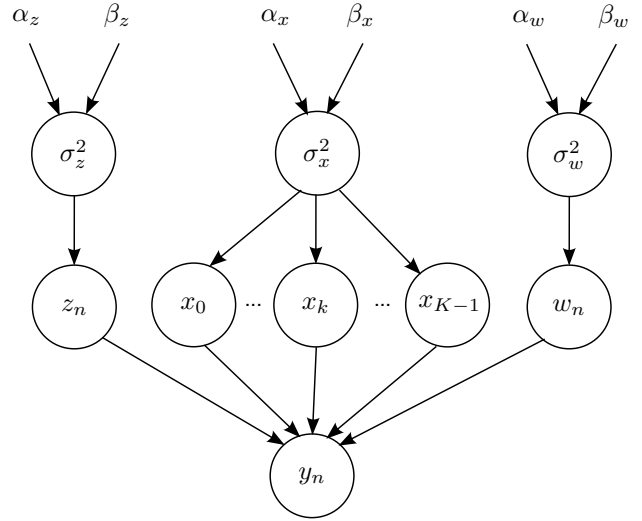}
\caption{Hierarchical Bayesian model of the problem. The observation $y_n$ depends on coefficients $x_0,\ldots,x_{K-1}$ and jitter and additive noise $z_n$ and $w_n$. The coefficients all depend on the signal variance $\sigma_x^2$, and the jitter and additive noise depend on $\sigma_z^2$ and $\sigma_w^2$, respectively. Each of these variances depend on hyperparameters $\alpha$ and $\beta$. In this model, circled nodes are random variables, and non-circled nodes are fixed parameters.}
\label{fig:hierbayesmodel}
\end{figure}

In this paper, we assume both the jitter and additive noise are random, independent of each other and the signal $x(t)$. Specifically, $z_n$ and $w_n$ are assumed to be iid zero-mean Gaussian, with variances equal to $\sigma_z^2$ and $\sigma_w^2$, respectively. In keeping with the Bayesian framework, we also choose a prior for the signal parameters. For convenience, we use an iid zero-mean Gaussian prior with variance $\sigma_x^2$ because the observation model is linear in the parameters. Rather than assuming these parameters (variances) to be known, we treat them as random variables and assign a conjugate prior to these parameters. Thus, $\sigma_z^2$, $\sigma_w^2$, and $\sigma_x^2$ are inverse Gamma distributed with hyperparameters $\{\alpha_z,\beta_z\}$, $\{\alpha_w,\beta_w\}$, and $\{\alpha_x,\beta_x\}$, respectively. These hyperparameters may be selected to be consistent with in-factory measurement of the noise variances or other information. The hierarchical Bayesian model is shown in Figure~\ref{fig:hierbayesmodel}.

To simplify notation, the probability density function (pdf) of $\mathbf{a}$ is written as $p(\mathbf{a})$, and the pdf of $\mathbf{b}$ conditioned on $\mathbf{a}$ is abbreviated as $p(\mathbf{b}\mid\mathbf{a})$ for random $\mathbf{a}$ and $p(\mathbf{b};\mathbf{a})$ for nonrandom $\mathbf{a}$. The subscripts usually included outside the parentheses will be written only when needed to avoid confusion. Expectations will follow the same convention.

The uniform distribution is written in this paper as $U(\text{set})$; for instance, $U([a,b])$ is a uniform distribution over the interval $[a,b]$, and $U(\{u : p(u) \geq c\})$ is a uniform distribution over the set $\{u : p(u) \geq c\}$. Writing $u \sim U(\text{set})$ means that $u$ is a sample generated from this distribution; analogous notation is used for the other distributions in this paper. The inverse Gamma distribution has the density function
\begin{equation}
\mathcal{IG}(s;\alpha,\beta) \defeq \frac{\beta^\alpha}{\Gamma(\alpha)}s^{-\alpha-1}e^{-\beta/s}.\label{eq:intro_igdist}
\end{equation}
The mean and variance of $s$ are
\begin{equation}
\mathbb{E}[s] = \frac{\beta}{\alpha-1},\quad \text{and}\quad \var(s) = \frac{\beta^2}{(\alpha-1)^2(\alpha-2)}.\label{eq:intro_igmeanvar}
\end{equation}
The density function of the $d$-dimensional normal distribution with mean $\boldsymbol{\mu}$ and covariance matrix $\boldsymbol{\Lambda}$ is written as
\begin{equation}
\mathcal{N}(\mathbf{a};\boldsymbol{\mu},\boldsymbol{\Lambda}) \defeq |2\pi\boldsymbol{\Lambda}|^{-1/2}\exp\{-\frac{1}{2}(\mathbf{a}-\boldsymbol{\mu})^T\boldsymbol{\Lambda}^{-1}(\mathbf{a}-\boldsymbol{\mu})\}.\label{eq:intro_normdist}
\end{equation}

When performing simulations, specific values are required for the $\alpha$'s and $\beta$'s. For the signal variance $\sigma_x^2$, consider an unbiased estimate of that variance from $K > 1$ observations generated from a standard normal distribution: $s_K = \frac{1}{K-1}\sum_{k=0}^{K-1} (x_k-\bar{x})^2$, where $\bar{x} = \frac{1}{K}\sum_{k=0}^{K-1} x_k$ is the sample mean. Then, we fit the inverse Gamma prior hyperparameters $\alpha_x$ and $\beta_x$ to the mean and variance of $s_K$ using~\eqref{eq:intro_igmeanvar}:
\begin{equation}
\frac{\beta_x}{\alpha_x-1} = \mathbb{E}[s_K] = 1;\ \frac{\beta_x^2}{(\alpha_x-1)^2(\alpha_x-2)} = \var(s_K) = \frac{2}{K-1}.
\end{equation}
Solving,
\begin{equation}
\alpha_x = \frac{K+3}{2};\ \beta_x = \frac{K+1}{2}.\label{eq:intro_alphabetax}
\end{equation}
Similarly for the zero-mean jitter and additive noise variances, given $N > 1$ observations and expected noise variances $\mathbb{E}[\sigma_z^2]$ and $\mathbb{E}[\sigma_w^2]$,
\begin{equation}
\alpha_z = \alpha_w = \frac{N+3}{2},\ \beta_z = \frac{N+1}{2}\mathbb{E}\left[\sigma_z^2\right],\ \text{and}\ \beta_w = \frac{N+1}{2}\mathbb{E}\left[\sigma_w^2\right].\label{eq:intro_alphabetazw}
\end{equation}
For the examples in this paper, we use the same $K$ and $N$ as for our signal; in practical applications, $K$ and $N$ are prior observations performed at a factory (for the noise variances) or elsewhere (for the signal variance).

The objective of the algorithm presented in this paper is to find the estimator $\mathbf{\hat{x}}$ that minimizes the mean squared error (MSE) $\mathbb{E}\left[\|\mathbf{\hat{x}(y)-x}\|_2^2\right]$, where the observations $\mathbf{y}$ are implicitly functions of $\mathbf{x}$. Unlike in the classical estimation framework, we have a prior on $\mathbf{x}$, which allows us to formulate the minimum mean squared error (MMSE) estimator $\mathbf{\hat{x}}_{\text{MMSE}}$ as the posterior expectation
\begin{equation}
\mathbf{\hat{x}}_{\text{MMSE}} \defeq \mathbb{E}[\mathbf{x}\mid\mathbf{y}].\label{eq:intro_mmseest}
\end{equation}
The posterior distribution $p(\mathbf{x}\mid\mathbf{y})$ depends on the likelihood function $p(\mathbf{y}\mid\mathbf{x})$, which can be expressed as in~\cite{WellerClassical} as a product of marginal likelihoods:
\begin{equation}
p(y_n\mid\mathbf{x}) = \iiint\mathcal{N}(y_n;\mathbf{h}_n^T(z_n)\mathbf{x},\sigma_w^2)\mathcal{N}(z_n;0,\sigma_z^2)\mathcal{IG}(\sigma_z^2;\alpha_z,\beta_z)\mathcal{IG}(\sigma_w^2;\alpha_w,\beta_w)\,dz_n\,d\sigma_z^2\,d\sigma_w^2.\label{eq:intro_pyxsep}
\end{equation}
As neither the likelihood nor posterior distribution has a simple closed form, the majority of this paper is devoted to approximating these functions using numerical and stochastic methods.

\subsection{Related Work}

Random jitter has been studied extensively throughout the early signal processing literature (see~\cite{Balakrishnan62}, \cite{Brown63}, and \cite{Liu65}). However, much of the effort in designing reconstruction algorithms was constrained to linear transformations of the observations. These papers also analyze the performance of such algorithms; for example, \cite{Liu65} proves that when the jitter is Gaussian and small enough, the MSE is approximately $\frac{1}{3}\Omega_B^2\sigma_z^2$, where the input PSD $S_{xx}(j\Omega) = \frac{1}{2\Omega_B}$ is flat. Due to the lack of attention to nonlinear post-processing, it is not readily apparent from the literature that these linear estimators are far from optimal. The effects of jitter on linear MMSE reconstruction of bandlimited signals are discussed in~\cite{Nordio09} and extended to the asymptotic $K,N \rightarrow \infty$ case and multidimensional signals in~\cite{Nordio10}.

More recently, \cite{Cox93} uses a second-order Taylor series approximation to perform weighted least-squares fitting of a jittered random signal. In~\cite{Tuncer07}, two post-processing methods are described for the case when the sample times are discrete (on a dense grid). Similar to the Gibbs sampler presented in this work, \cite{Andrieu96} uses a Metropolis-Hastings Markov chain Monte Carlo (MCMC) algorithm to estimate the jitter and jitter variance from a sequence of samples. Also, a maximum a posteriori (MAP)-based estimator is proposed in~\cite{Zhang07} to mitigate read-in and write-out jitter in data storage devices. Finally, a Gibbs sampler is developed in~\cite{Tan08} to estimate the coefficients and locations of finite rate of innovation signals from noisy samples.

Preliminary versions of the algorithms and results presented in this work are also discussed, with further background material and references, in~\cite{WellerThesis}.

\subsection{Outline}

In Section~\ref{sec:background}, numerical quadrature is revisited and Gibbs sampling and slice sampling are reviewed. The linear MMSE estimator is discussed in Section~\ref{sec:linear}. In Section~\ref{sec:nonlinear}, the Gibbs sampler approximation to the Bayes MMSE estimator is derived, and slice sampling is used in the implementation. All these estimators, as well as the EM algorithm from~\cite{WellerClassical} approximating the ML estimator, are analyzed and compared via simulations in Section~\ref{sec:simresults}. Conclusions based on these simulations, as well as ideas for future research directions, are discussed in Section~\ref{sec:conclusion}.

\section{Background}\label{sec:background}

In general, the likelihood function in the introduction is described in terms of an integration without a closed form. Fortunately, numerical methods such as Gauss quadrature, which approximates the integration in question with a weighted sum of the integrand evaluated at different locations (abscissas), are relatively accurate and efficient. A more detailed description of Gauss quadrature can be found in the background section of~\cite{WellerClassical}, or in~\cite{Kythe05} or~\cite{Davis84}. This paper discusses using Gauss--Laguerre quadrature to approximate integration with respect to $\sigma_z^2$ and $\sigma_w^2$.

However, simply being able to evaluate (approximately) the likelihood function is insufficient to approximate the Bayes MMSE estimator. To approximate the expectation in~\eqref{eq:intro_mmseest}, we propose using a Monte Carlo statistical method combining Gibbs sampling and slice sampling. Gibbs sampling and slice sampling are discussed below.

\subsection{Numerical Integration}

For integrals of the form $\int_{-\infty}^\infty f(x)\mathcal{N}(x;\mu,\sigma^2)\,dx$, techniques such as Gauss--Legendre and Gauss--Hermite quadrature, Romberg's method, and Simpson's rule, are described in~\cite{WellerClassical}. Similarly, Gauss--Laguerre quadrature can approximate integrals of the form $\int_0^\infty f(x)x^ae^{-x}\,dx$. The abscissas and weights for Gauss--Laguerre quadrature can be computed using the eigenvalue-based method derived in~\cite{Golub69}.

Let $x_j$ and $w_j$ be the abscissas and weights for the Gauss--Laguerre quadrature rule of length $J$. Then, we can integrate against the pdf of the inverse Gamma distribution by observing,
\begin{equation}
\begin{split}
\int_0^\infty f(x)\mathcal{IG}(x;\alpha,\beta)\,dx &= \int_0^\infty \frac{\beta^\alpha}{\Gamma(\alpha)}f(x)x^{-(\alpha+1)}e^{-\beta/x}\,dx\\
&= \int_0^\infty \frac{\beta^\alpha}{\Gamma(\alpha)}f\left(\frac{\beta}{y}\right)\left(\frac{y}{\beta}\right)^{\alpha+1}\frac{\beta}{y^2}e^{-y}\,dy\\
&= \int_0^\infty \frac{1}{\Gamma(\alpha)}f\left(\frac{\beta}{y}\right)y^{\alpha-1}e^{-y}\,dy\\
&\approx \sum_{j=1}^J w_j'f(x_j'),\end{split}\label{eq:bg_genlagrule}
\end{equation}
where $x_j' = \beta/x_j$, and $w_j' = w_j/\Gamma(\alpha)$. The substitutions $x = \beta/y$ and $dx = \beta/y^2\,dy$ are made in the second step of the derivation.

\begin{figure}
\centering
\subfloat[][$K = 10$, $M = 4$, $\mathbb{E}\lbr\sigma_z^2\rbr = 0.75^2$, $\mathbb{E}\lbr\sigma_w^2\rbr = 0.1^2$, $J_1 = J_2 = 9$, $J_3 = 129$, $n = 19$.]{\includegraphics[width=3in]{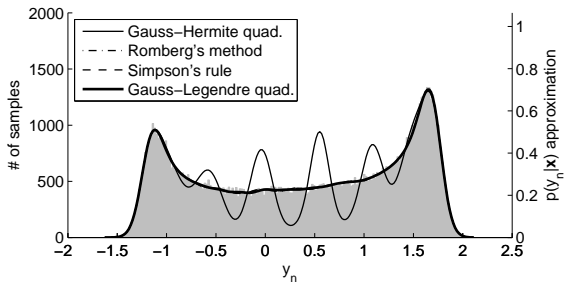}\label{fig:hermitevalid-1}}\\
\subfloat[][$K = 10$, $M = 4$, $\mathbb{E}\lbr\sigma_z^2\rbr = \mathbb{E}\lbr\sigma_w^2\rbr = 0.01^2$, $J_1 = J_2 = 9$, $J_3 = 129$, $n = 18$.]{\includegraphics[width=3in]{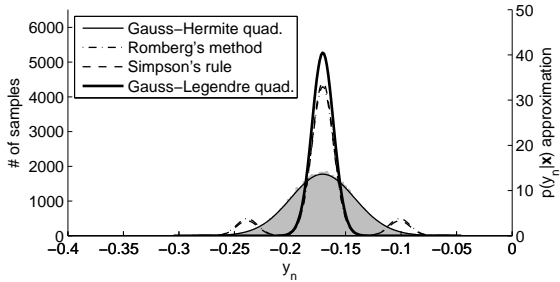}\label{fig:hermitevalid-3}}\\
\caption{Quadrature approximations are compared to histograms for $p(y_n\mid\mathbf{x})$ for different expected values of $\sigma_z^2$ and $\sigma_w^2$ ($\alpha$'s and $\beta$'s are computed according to~\eqref{eq:intro_alphabetazw}). The quadrature approximations are computed for a dense grid of $200$ values of $y_n$, and the histograms are generated from $100\,000$ samples of $y_n$, computed from samples of $\sigma_z^2$, $\sigma_w^2$, $z_n$, and $w_n$ according to~\eqref{eq:intro_obsmodel}. The multimodal case~\protect\subref{fig:hermitevalid-1} favors Gauss--Legendre quadrature; case~\protect\subref{fig:hermitevalid-3} favors Gauss--Hermite quadrature; the worst-case $n$ is shown in each. The legend refers to the quadrature method used for the integral over $z_n$; Gauss--Laguerre quadrature is used for the integrals with respect to $\sigma_z^2$ and $\sigma_w^2$.\label{fig:hermitevalid}}
\end{figure}

Utilizing a combination of Gauss--Laguerre quadrature and either Gauss--Hermite quadrature or Gauss--Legendre quadrature, we can approximate the likelihood function $p(y_n\mid x)$ using the integral in~\eqref{eq:intro_pyxsep}. In particular,
\begin{equation}
p(y_n\mid x) \approx \sum_{j_1=1}^{J_1}\sum_{j_2}^{J_2}\sum_{j_3}^{J_3} w_{j_1}w_{j_2}w_{j_3}\mathcal{N}(y_n;\mathbf{h}_n^T(z_{j_3})\mathbf{x},{\sigma_w^2}_{j_1}).\label{eq:bg_lapproxquad}
\end{equation}
In this equation, the innermost quadrature (over $z_n$) depends on the value of $\sigma_z^2$, so the values of $z_{j_3}$ depend on ${\sigma_z^2}_{j_2}$. Since the total number of operations scales exponentially with the number of variables being integrated, we seek to minimize the choices of $J_1$, $J_2$, and $J_3$ for this three-dimensional summation. To explore the accuracy of this approximation as a function of $J_1$ and $J_2$ (we use $J_3 = 129$ from~\cite{WellerClassical}), the quadrature is performed over a dense grid of values of $y_n$ and the results are compared to a histogram generated empirically, by fixing $\mathbf{x}$ to a randomly chosen vector, generating many samples of $\sigma_z^2$, $\sigma_w^2$, $z_n$, and $w_n$ from their respective prior distributions, and computing the samples $y_n$ using~\eqref{eq:intro_obsmodel}. Comparisons for unimodal and multimodal $p(y_n;\mathbf{x})$ are shown in Figure~\ref{fig:hermitevalid}. Based on these comparisons, we choose Gauss--Laguerre quadrature with $J_1 = J_2 = 9$ to integrate with respect to $\sigma_w^2$ and $\sigma_z^2$. This is combined with Gauss--Hermite quadrature with $J_3 = 129$ when $\mathbb{E}[\sigma_z^2]$ is small ($< 0.01$) and Gauss--Legendre quadrature with $J_3 = 129$ when $\mathbb{E}[\sigma_z^2] > 0.01$. This hybrid quadrature also is used when computing the expectations in Section~\ref{sec:linear} and in the appendix.

\subsection{Gibbs Sampling}

The Gibbs sampler is a Markov chain Monte Carlo method developed in~\cite{Geman84}. Details about the Gibbs sampler and its many variants, including Metropolis-within-Gibbs sampling, can be found in~\cite{Robert04}. When implementing the Gibbs sampler, one must consider both the number of iterations until the Markov chain has approximately converged to its stationary distribution (the ``burn-in time'') and the number of samples that should be taken after convergence to compute the MMSE estimate. According to~\cite{Smith93}, separating highly correlated variables slows convergence of the Gibbs sampler. The number of iterations after convergence is connected to both correlation between successive samples and the variance of the random variables distributed according to the stationary distribution.

To monitor convergence, heuristics such as the potential scale reduction factor (PSRF) and the inter-chain and intra-chain variances are developed in~\cite{Gelman92,Brooks98}. Consider $C$ instances (chains) of the Gibbs sampler running simultaneously. Define the vector $\mathbf{a}_{c,i}$ to be the combined vector of all the samples for the $c$th chain at the $i$th iteration. For chain $c$, the average is $\mathbf{\bar{a}}_c = \frac{1}{i}\sum_{j=1}^i \mathbf{a}_{c,i}$. Across all chains, the average is $\mathbf{\bar{\bar{a}}} = \frac{1}{C}\sum_{c=1}^C \mathbf{\bar{a}}_c$. Then, following the multivariate extension to the potential scale reduction factor (PSRF) derived in~\cite{Brooks98}, define the intra-chain covariance
\begin{equation}
\mathbf{W}_i \defeq \frac{1}{(i-1)C} \sum_{c=1}^C \sum_{j=1}^i (\mathbf{a}_{c,i}-\mathbf{\bar{a}}_c)(\mathbf{a}_{c,i}-\mathbf{\bar{a}}_c)^T,\label{eq:nl_covintrachain}
\end{equation}
and the inter-chain covariance
\begin{equation}
\mathbf{B}_i \defeq \frac{1}{i-1} \sum_{j=1}^i (\mathbf{\bar{a}}_c-\mathbf{\bar{\bar{a}}})(\mathbf{\bar{a}}_c-\mathbf{\bar{\bar{a}}})^T.\label{eq:nl_covinterchain}
\end{equation}
The posterior variance $\mathbf{\hat{V}}_i \defeq \frac{i-1}{i}\mathbf{W}_i+\frac{C+1}{C}\mathbf{B}_i$, and the PSRF $\hat{R}^p \defeq \frac{i-1}{i}+\frac{C+1}{C}\|\mathbf{W}_i^{-1}\mathbf{B}_i\|_2$, where $\|\cdot\|_2$ is the induced matrix $2$-norm. Then, the Gibbs sampler's Markov chain has converged when $\hat{R}^p = 1$, and $\hat{\mathbf{V}}$ stabilizes. To measure the change in $\hat{\mathbf{V}}$, we compute $\|\hat{\mathbf{V}}\|_2^{1/2}$.

\subsection{Slice Sampling}

\begin{figure}[!t]
\centering
\psfrag{x}[][]{$x$}
\psfrag{y}[][]{$y$}
\psfrag{xi1}[][]{$x^{(i-1)}$}
\psfrag{xi}[][]{$x^{(i)}$}
\psfrag{yi}[][]{$y^{(i)}$}
\psfrag{pxi1}[][]{$p(x^{(i-1)})$}
\psfrag{pxi}[][]{$x^{(i)} \sim U(\{x:p(x)\geq y^{(i)}\})$}
\psfrag{pyi}[][]{$y^{(i)} \sim U([0,p(x^{(i-1)})])$}
\psfrag{a}[][]{$(a)$}
\psfrag{b}[][]{$(b)$}
\includegraphics[width=3.45in]{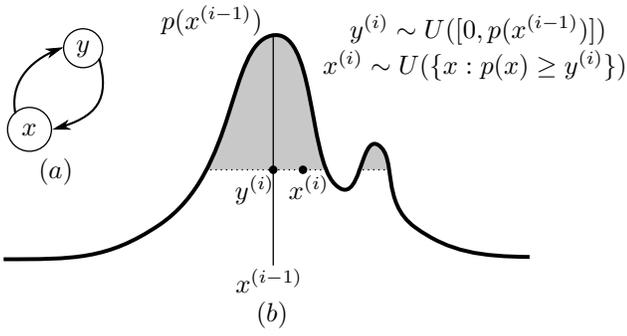}
\caption{Slice sampling of $p(x)$ illustrated: (a) Sampling is performed by traversing a Markov chain to approximate $p(x)$, the stationary distribution. Each iteration consists of (b) uniformly choosing a slice $\{x : p(x) \geq y\}$ and uniformly picking a new sample $x$ from that slice.}
\label{fig:slicesampling}
\end{figure}
Slice sampling is a Markov chain Monte Carlo method described in~\cite{Neal03} for generating samples from a distribution by instead sampling uniformly from the subgraph of the pdf and framing this sampling procedure as a two-stage Gibbs sampler, depicted in Figure~\ref{fig:slicesampling}.

The difficulty of slice sampling is in representing and sampling from the slice. In this problem, we show that any given slice is bounded, and therefore, an interval containing the slice can be constructed, and the ``shrinkage'' method described in~\cite{Neal03} can be used. The shrinkage method is an accept-reject method, where given an interval $[L,R]$ containing part (or all) of the slice, a sample is generated uniformly from the interval and accepted if the sample is inside the slice. If the sample is rejected, the interval shrinks to use the rejected sample as a new endpoint. Several variants, including shrinking to the midpoint of the interval instead of or in addition to the rejected sample, are also described in the rejoinder at the end of~\cite{Neal03}. These variants are compared in the context of the jitter mitigation problem in Section~\ref{sec:nonlinear}.

\section{Linear Bayesian Estimation}\label{sec:linear}

When block post-processing the samples, the linear Bayesian estimator with minimum MSE is called the linear MMSE (abbreviated LMMSE) estimator. The general form of the linear MMSE estimator is given in~\cite{Kay93}. For estimating the random signal coefficients $\mathbf{x}$ using the hierarchical Bayesian model in Section~\ref{sec:intro},
\begin{align}
\boldsymbol{\Lambda_{xy}} &= \frac{\beta_x}{\alpha_x-1}\mathbb{E}[\mathbf{H(z)}]^T,\\ \boldsymbol{\Lambda_y} &= \frac{\beta_x}{\alpha_x-1}\mathbf{E}[\mathbf{H(z)}\mathbf{H(z)}^T]+\frac{\beta_w}{\alpha_w-1}\mathbf{I},
\end{align}
and $\boldsymbol{\mu_y} = \boldsymbol{\mu_x} = \mathbf{0}$. The LMMSE estimator for random jitter is
\begin{equation}
\mathbf{\hat{x}}_{\text{LMMSE}}(\mathbf{y}) = \mathbb{E}[\mathbf{H(z)}]^T\left(\mathbb{E}[\mathbf{H(z)}\mathbf{H(z)}^T]+\frac{\beta_w(\alpha_x-1)}{\beta_x(\alpha_w-1)}\mathbf{I}\right)^{-1}\mathbf{y}.\label{eq:lin_lmmseest}
\end{equation}
The expectations in~\eqref{eq:lin_lmmseest} can be computed off-line using Gauss quadrature. The error covariance of the LMMSE estimator is also derived in~\cite{Kay93}; for this problem,
\begin{equation}
\boldsymbol{\Lambda}_{\text{LMMSE}} = \frac{\beta_x}{\alpha_x-1}\left(\mathbf{I} - \mathbb{E}[\mathbf{H(z)}]^T\left(\mathbb{E}[\mathbf{H(z)}\mathbf{H(z)}^T]+\frac{\beta_w(\alpha_x-1)}{\beta_x(\alpha_w-1)}\mathbf{I}\right)^{-1}\mathbb{E}[\mathbf{H(z)}]\right).\label{eq:lin_lmmseerrcov}
\end{equation}

When no jitter is assumed, the LMMSE estimator simplifies to
\begin{equation}
\mathbf{\hat{x}}_{\text{LMMSE}\mid\mathbf{z}=\mathbf{0}}(\mathbf{y}) = \mathbf{H(0)}^T\left(\mathbf{H(0)}\mathbf{H(0)}^T+\frac{\beta_w(\alpha_x-1)}{\beta_x(\alpha_w-1)}\mathbf{I}\right)^{-1}\mathbf{y}.\label{eq:lin_lmmseestnoz}
\end{equation}
This linear estimator is the best linear transformation of the data that can be performed in the absence of jitter. Hence, the no-jitter LMMSE estimator is the baseline estimator against which the nonlinear Bayesian estimators derived later are measured. The error covariance of this estimator is
\begin{align}
\boldsymbol{\Lambda}_{\text{LMMSE}\mid\mathbf{z}=\mathbf{0}} &= \sigma_x^2\left(\mathbf{I} - \mathbf{H(0)}^T\left(\mathbf{H(0)}\mathbf{H(0)}^T+\frac{\beta_w(\alpha_x-1)}{\beta_x(\alpha_w-1)}\mathbf{I}\right)^{-1}\mathbf{H(0)}\right).\label{eq:lin_lmmsenozerrcov}
\end{align}

\section{Nonlinear Bayesian Estimation}\label{sec:nonlinear}

To improve upon the LMMSE estimator, we expand our consideration to nonlinear functions of the data. The Bayes MMSE estimator, in its general form in~\eqref{eq:intro_mmseest}, is the nonlinear function that minimizes the MSE. However, since the posterior density function for this problem does not have a closed form, this estimator can be difficult to compute. Since we are interested in the mean of the posterior pdf, finding the Bayes MMSE estimator is an obvious application of Monte Carlo statistical methods, especially the Gibbs sampler described in Section~\ref{sec:background}.

We propose using Gibbs sampling to produce a sequence of samples for the random parameters we wish to find, via traversing a Markov chain to its steady-state distribution, and average the samples to approximate the estimator. To this end, samples of $\mathbf{z}$, $\mathbf{x}$, $\sigma_x^2$, $\sigma_w^2$, and $\sigma_z^2$ are generated according to their full conditional distributions (i.e. the distribution of one random variable given all the others). To generate samples of $\mathbf{z}$, we apply slice sampling.

\subsection{Generating $z_n$ using Slice Sampling}
Consider generating samples $z_n$ from the distribution $p(\cdot\mid \mathbf{z}_{\backslash n},\mathbf{x},\sigma_x^2,\sigma_w^2,\sigma_z^2,\mathbf{y})$, where $\mathbf{z}_{\backslash n}$ is the random vector of all the jitter variables except $z_n$. Using Bayes rule and the independence of $z_n$ and $w_n$,
\begin{equation}
\begin{split}
p(z_n\mid \mathbf{z}_{\backslash n},\mathbf{x},\sigma_x^2,\sigma_w^2,\sigma_z^2,\mathbf{y}) &= \frac{p(\mathbf{y}\mid \mathbf{z},\mathbf{x},\sigma_x^2,\sigma_w^2,\sigma_z^2)p(\mathbf{z},\sigma_z^2)p(\mathbf{x}\mid\sigma_x^2)p(\sigma_x^2)p(\sigma_w^2)}{p(\mathbf{z}_{\backslash n},\mathbf{x},\mathbf{y},\sigma_x^2,\sigma_w^2,\sigma_z^2)}\\
&\propto \mathcal{N}(y_n;\mathbf{h}_n^T(z_n)\mathbf{x},\sigma_w^2)\mathcal{N}(z_n;0,\sigma_z^2).\end{split}\label{eq:nl_pzncond}
\end{equation}

%
Slice sampling is used for generating realizations of $z_n$ since no tightly enveloping proposal density or other tuning is necessary; the ability to evaluate an unnormalized form of the target distribution is sufficient. Each iteration of slice sampling consists of two uniform sampling problems:
\begin{enumerate}
\item Choose a slice $u$ uniformly from $[0,\tilde{p}(z_n^{(i)}\mid \mathbf{y},\mathbf{x},\sigma_x^2,\sigma_w^2,\sigma_z^2)]$, where $\tilde{p}(z_n^{(i)}\mid \mathbf{y},\mathbf{x},\sigma_x^2,\sigma_w^2,\sigma_z^2)$ is the unnormalized full conditional density function in~\eqref{eq:nl_pzncond}.
\item Sample $z_n^{(i+1)}$ uniformly from the slice $S \defeq \{z_n : \tilde{p}(z_n\mid \mathbf{y},\mathbf{x},\sigma_x^2,\sigma_w^2,\sigma_z^2) \geq u\}$.
\end{enumerate}
The first step is trivial, since we are sampling from a single interval. The second step is more difficult. However, since $u \leq \tilde{p}(z_n\mid \mathbf{y},\mathbf{x},\sigma_x^2,\sigma_w^2,\sigma_z^2)$ for all $z_n$ in the slice,
\begin{equation}
\begin{split}
\log u &\leq -\frac{(y_n - \mathbf{h}_n^T(z_n)\mathbf{x})^{2}}{2\sigma_w^2} - \frac{z_n^2}{2\sigma_z^2} - \log (2\pi\sigma_z\sigma_w)\\
&\leq - \frac{z_n^2}{2\sigma_z^2} - \log (2\pi\sigma_z\sigma_w).\end{split}\label{eq:nl_sliceconstraint}
\end{equation}
Solving for $z_n$, the range of possible $z_n$ is bounded:
\begin{equation}
|z_n| \leq \sigma_z\sqrt{-2\log u - 2\log(2\pi\sigma_w\sigma_z)}.\label{eq:nl_slicezrange}
\end{equation}
Using these extreme points for the initial interval containing the slice, and the ``shrinkage'' method specified in~\cite{Neal03} to sample from the slice by repeatedly shrinking the interval, slice sampling becomes a relatively efficient method. The ``shrinkage'' method decreases the size of the interval exponentially fast, on average. To see this, consider one iteration of shrinkage, where the initial point $x_0$ from the previous step of slice sampling lies in the interval $[L,R]$. This initial point is guaranteed to be in the slice by construction. The expected size of the new interval $[L',R']$, from choosing a new point $x'$, is
\begin{equation}
\begin{split}
\mathbb{E}[R'-L'\mid R,L,x_0] &= \frac{1}{R-L}\left[\int_L^{x_0} (R - x')\,dx' + \int_{x_0}^R (x'-L)\,dx'\right]\\
&= \frac{R^2-2RL+L^2}{2(R-L)} + \frac{x_0(R+L-x_0) - RL}{R-L}\\
&= \frac{R-L}{2} + \frac{x_0(R+L-x_0) - RL}{R-L}.\end{split}\label{eq:nl_sliceshrinkwidth}
\end{equation}
This expectation is quadratic in $x_0$, so the maximum occurs at the extreme point $x_0 = (R+L)/2$. The maximum value is
\begin{equation}
\begin{split}
\max_{x_0} \mathbb{E}[R'-L'\mid R,L,x_0] &= \frac{R-L}{2} + \frac{((R+L)/2)(R+L-(R+L)/2) - RL}{R-L}\\
&= \frac{R-L}{2} + \frac{(R+L)^2/4 - RL}{R-L} = \frac{3}{4}(R-L).\end{split}\label{eq:nl_sliceshrinkwidthmax}
\end{equation}
Concavity implies that the minima are at the two endpoints $x_0 = L$ and $x_0 = R$. In both cases, the expected size of the interval is $(R-L)/2$. Therefore,
\begin{equation}
\frac{1}{2}(R-L) \leq \mathbb{E}[R'-L'\mid R,L,x_0] \leq \frac{3}{4}(R-L),\label{eq:nl_sliceshrinkwidthrange}
\end{equation}
which implies that at worst, the size of the interval shrinks to $3/4$ its previous size per iteration, on average. Then, given the initial interval $[L_0,R_0]$ and previous point $x_0$, the expected size of the interval $[L_I,R_I]$ after $I$ iterations of the shrinkage algorithm is
\begin{equation}
\begin{split}
\mathbb{E}[R_I-L_I\mid R_0,L_0,x_0] &= \mathbb{E}[\mathbb{E}[R_I-L_I\mid R_0,L_0,\ldots,R_{I-1},L_{I-1},x_0] \mid R_0, L_0, x_0]\\
&\leq \left(\frac{3}{4}\right)^I(R_0-L_0).\end{split}\label{eq:nl_sliceshrinkiterwidth}
\end{equation}
If the target distribution $p(x)$ is continuous, the algorithm is guaranteed to terminate once the search interval is small enough. Since the interval size shrinks exponentially fast, on average, the number of ``shrinkage'' iterations is approximately proportional to the log of the fraction of the initial interval contained in the slice.

\begin{figure*}
\centering
\psfrag{ptilde}[][]{$\scriptstyle\log\tilde{p}(z_n\mid\cdots)$}
\psfrag{zni1}[t][t][0.75]{$z_n^{(i-1)}$}
\psfrag{Lz}[t][t][0.75]{$L\leftarrow z$}
\psfrag{Rz}[t][t][0.75]{$R\leftarrow z$}
\psfrag{Lmidpt}[t][t][0.75]{$L\leftarrow \text{midpt.}$}
\psfrag{Rmidpt}[t][t][0.75]{$R\leftarrow \text{midpt.}$}
\psfrag{zniz}[t][t][0.75]{$z_n^{(i)}\leftarrow z$}
\subfloat[][$K = 10$, $M = 16$, $\sigma_z = 0.1$, $\sigma_w = 0.05$, $n = 80$, original (left) and thresholding-based (right) methods.]{\includegraphics[width=3in]{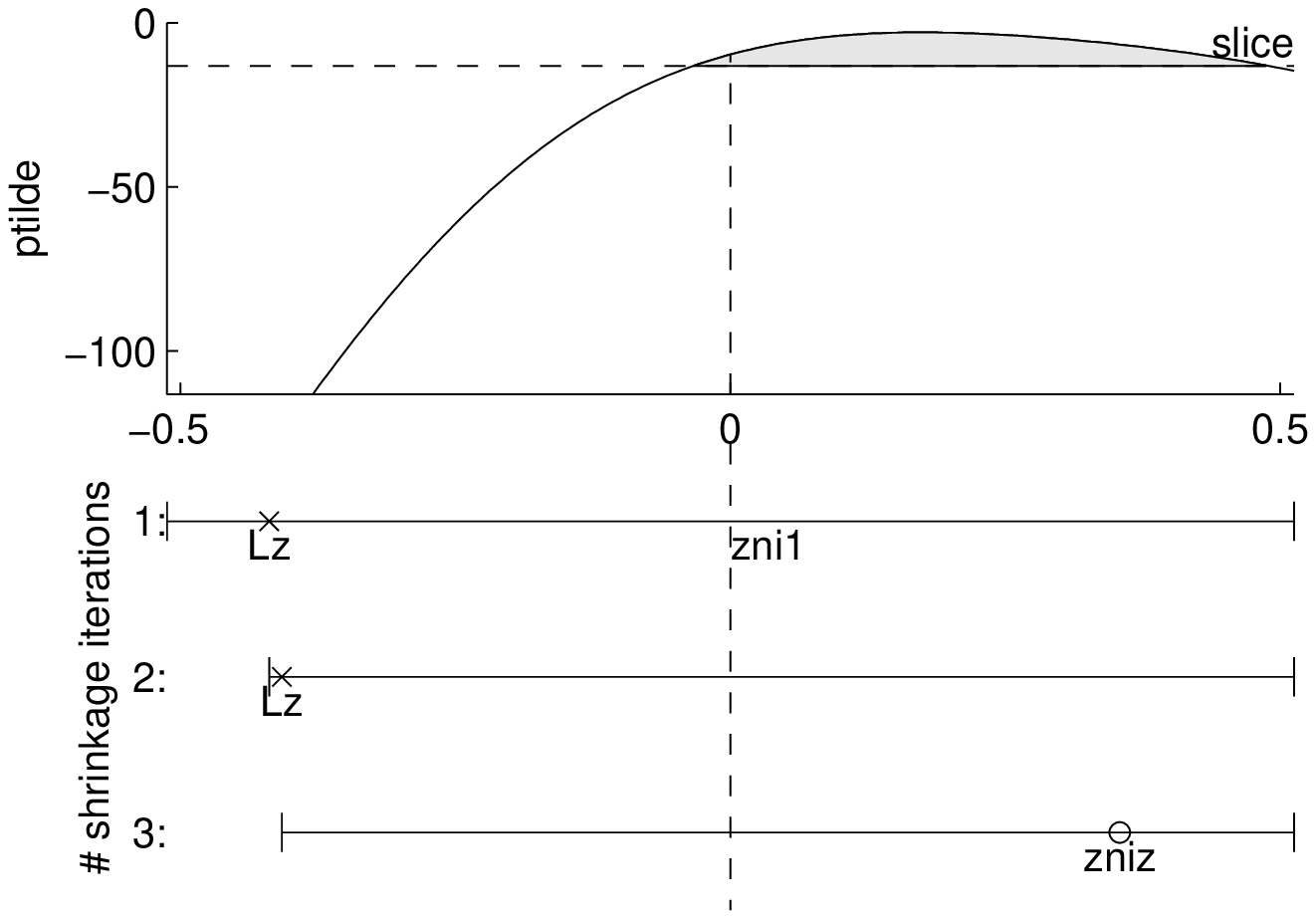}\quad\includegraphics[width=3in]{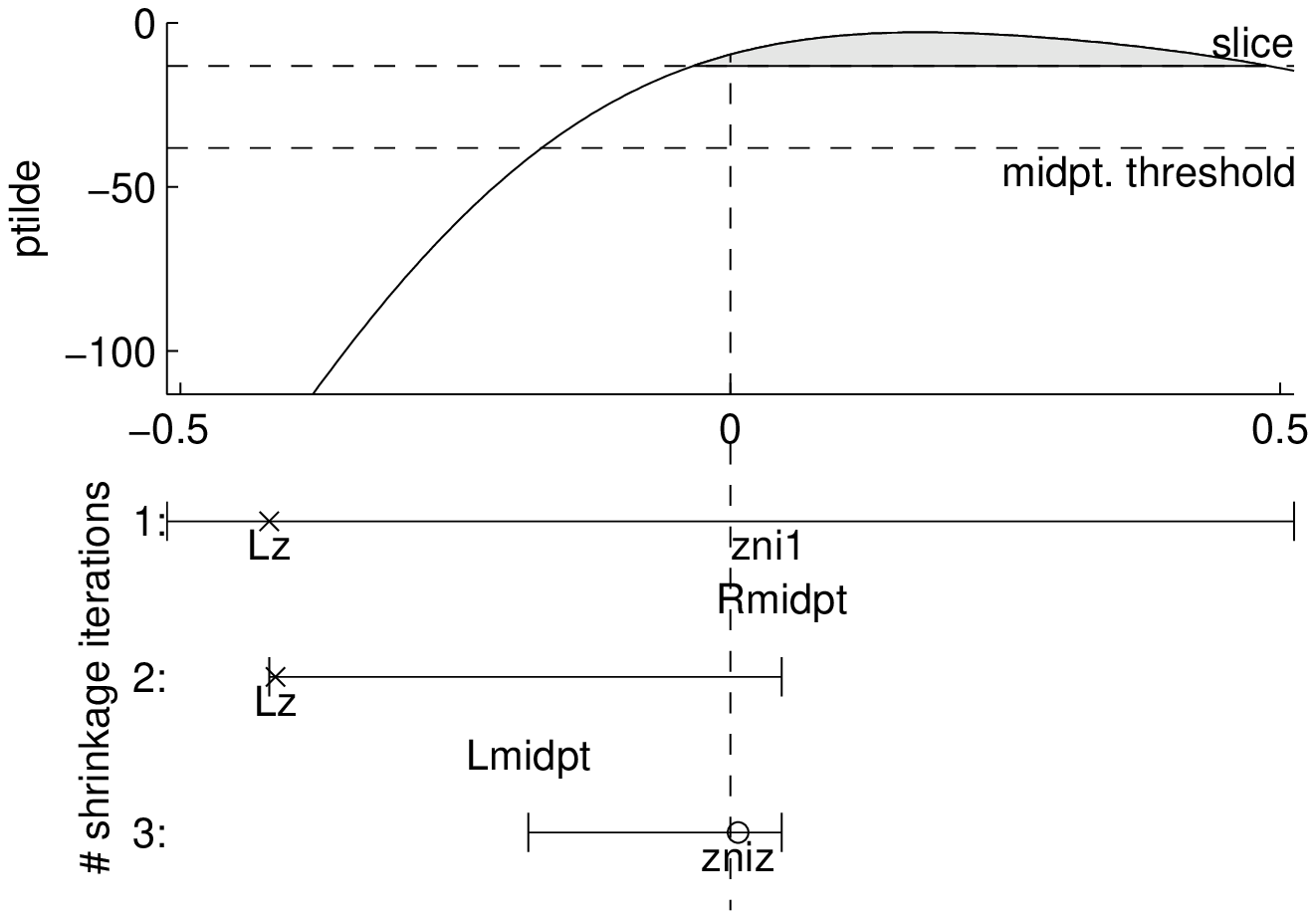}\label{fig:shrinkcompplots-1}}\\
\subfloat[][$K = 10$, $M = 4$, $\sigma_z = 0.5$, $\sigma_w = 0.075$, $n = 20$, original (left) and thresholding-based (right) methods.]{\includegraphics[width=3in]{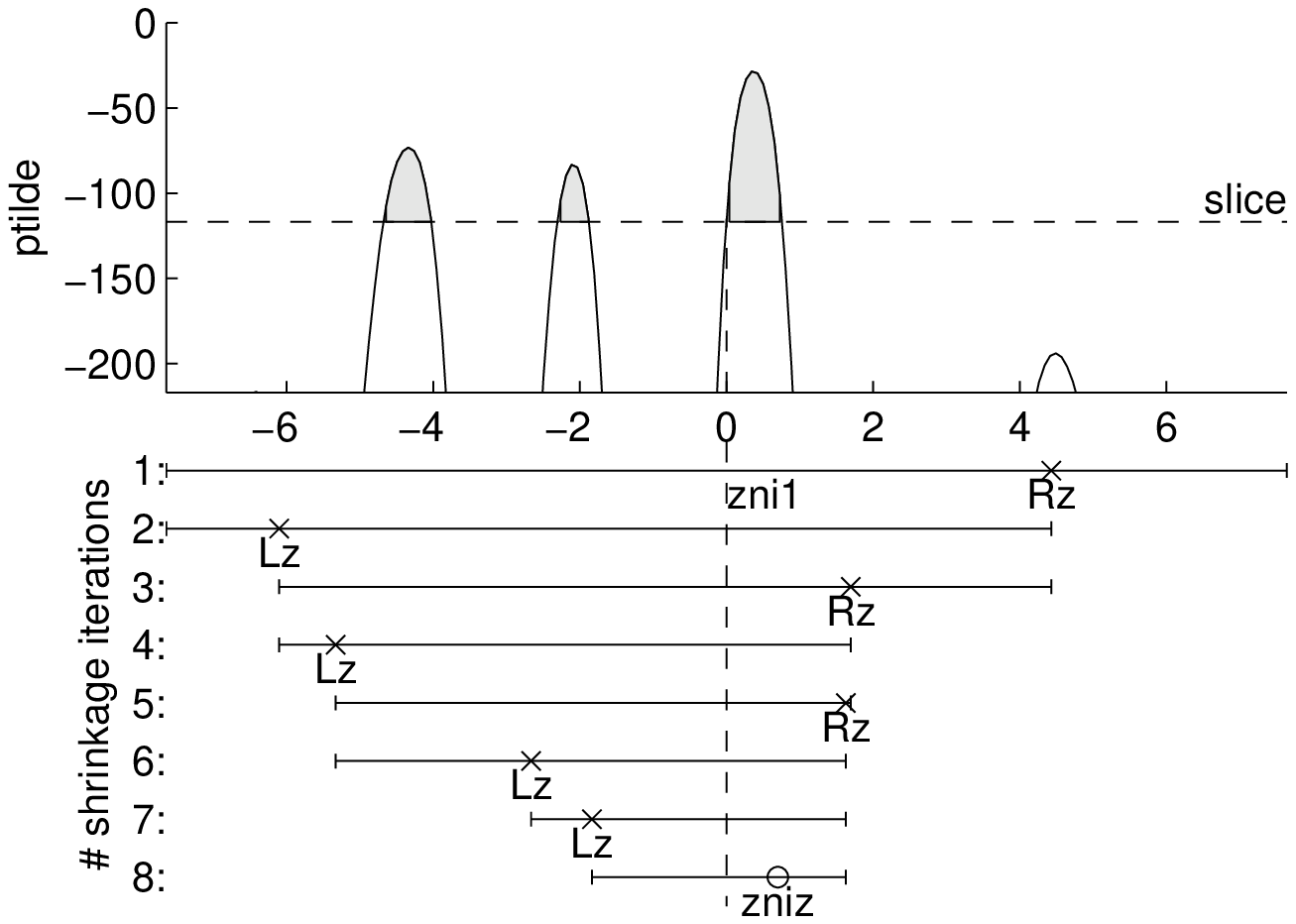}\quad\includegraphics[width=3in]{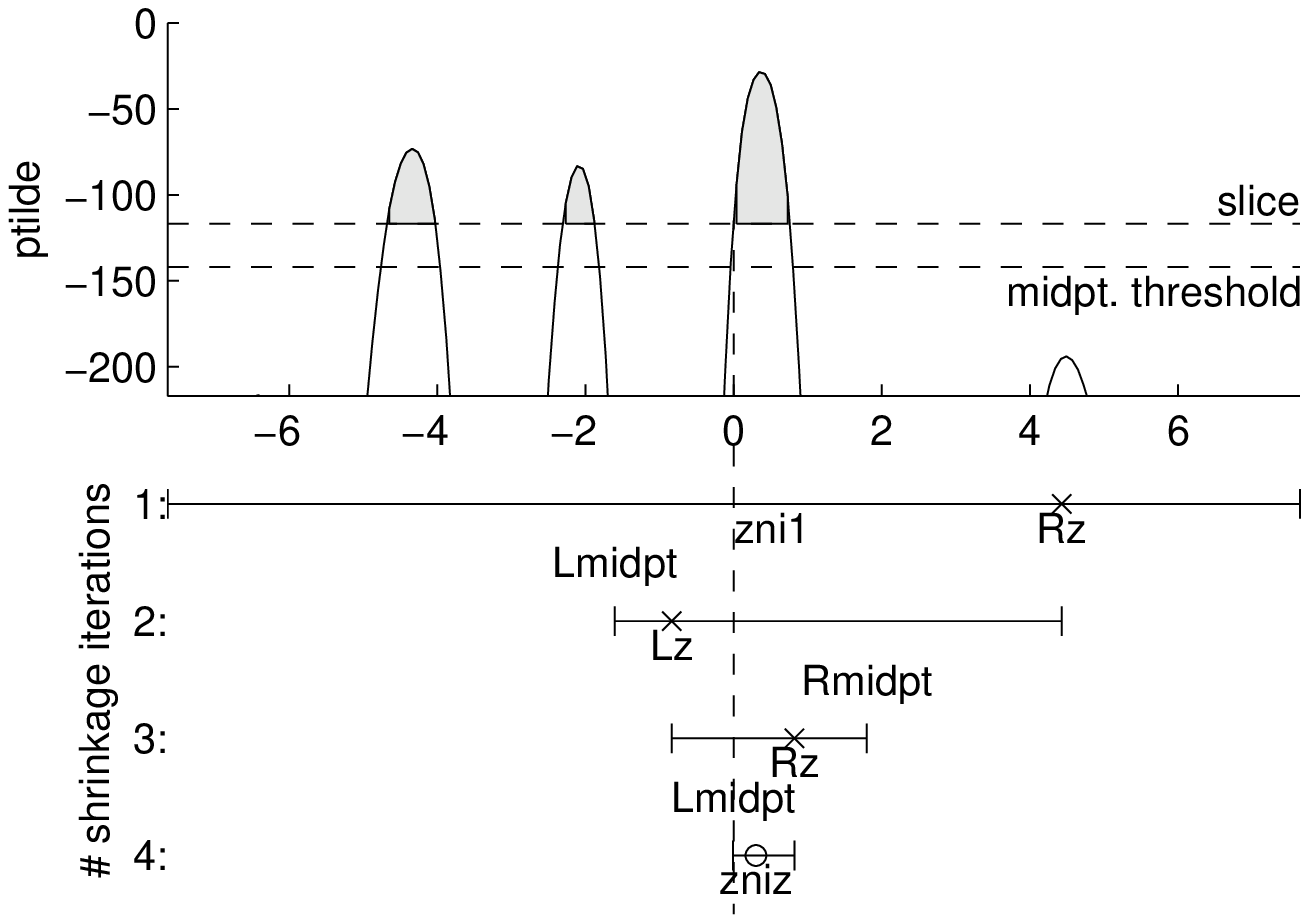}\label{fig:shrinkcompplots-2}}
\caption{Comparisons between the original shrinkage method and the modified thresholding-based shrinkage method for unimodal~\protect\subref{fig:shrinkcompplots-1} and multimodal~\protect\subref{fig:shrinkcompplots-2} posterior distributions. The unnormalized distribution $\tilde{p}(z_n^{(i-1)}\mid \mathbf{x},\sigma_x^2,\sigma_z^2,\sigma_w^2,y_n)$ is evaluated for the previous Gibbs sampler iteration's $z_n^{(i-1)}$ and the slice level shown is selected uniformly from $[0,\tilde{p}(z_n^{(i-1)}\mid\cdots)]$. The shrinkage methods proceed according to Algorithm~\ref{alg:nl_znslicesamp}. The midpoint threshold shown corresponds to $\tau = 25$. Rejected samples are marked with a ``$\times$'', and the final accepted sample is marked with a ``$\circ$''. In both cases, the thresholding-based method reduces the size of the interval more quickly than the original method. Especially in the unimodal case, the accepted sample $z_n^{(i)}$ is much closer to the previous iterate $z_n^{(i-1)}$ than would otherwise be expected from the size of the slice.}
\label{fig:shrinkcompplots}
\end{figure*}
In the rejoinder at the end of~\cite{Neal03}, an alternative binary search-like midpoint shrinkage algorithm is proposed that can converge faster on the slice than the original shrinkage algorithm, at the cost of increasing correlation between successive samples, which reduces the overall Gibbs sampler convergence speed. In an effort to mitigate the increased correlation, a hybrid method is proposed in~\cite{Neal03} that always shrinks to the rejected sample, then shrinks to the midpoint of the remaining interval only if the probability of the rejected sample is sufficiently small (the threshold) relative to the slice. These algorithms are applied to both unimodal and multimodal posterior distributions $p(z_n\mid\mathbf{x},\mathbf{z},\mathbf{y},\sigma_x^2,\sigma_z^2,\sigma_w^2)$ in Figure~\ref{fig:shrinkcompplots}.

\begin{figure}[!t]
\centering
\subfloat[][$K = 10$, $M = 4$, $\mathbb{E}\lbr\sigma_z^2\rbr = 0.5^2$, $\mathbb{E}\lbr\sigma_w^2\rbr = 0.1^2$.]{\includegraphics[width=3.2in]{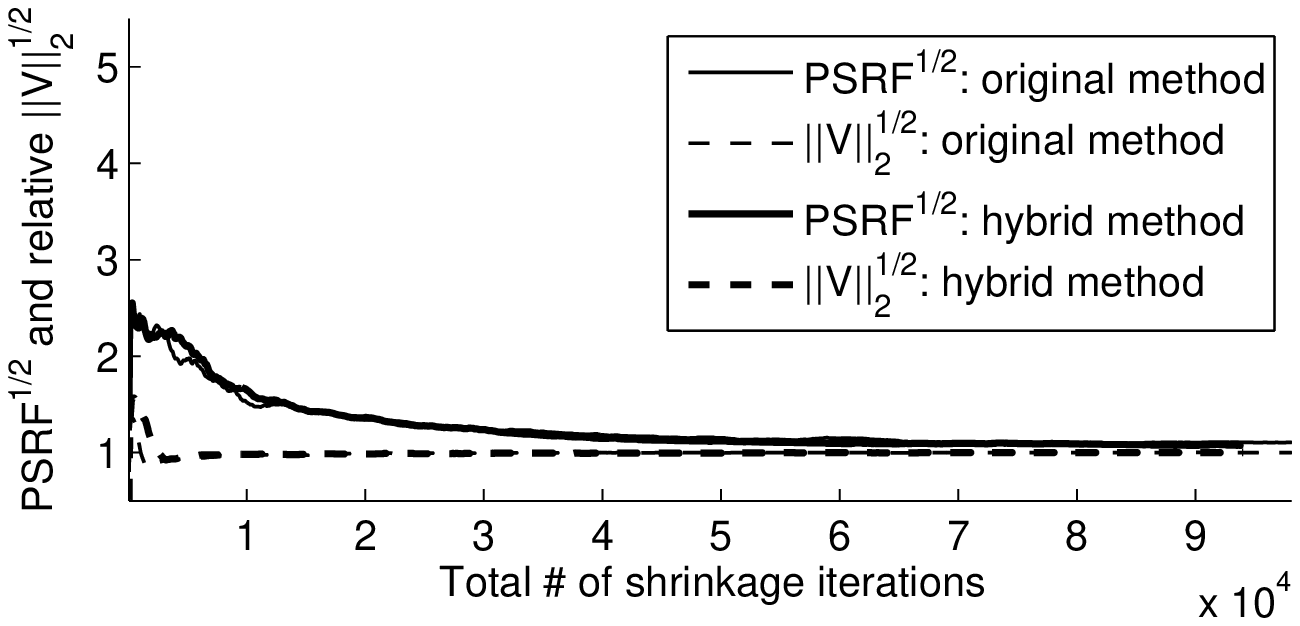}\label{fig:shrinkcomp-1}}\\
\subfloat[][$K = 10$, $M = 16$, $\mathbb{E}\lbr\sigma_z^2\rbr = 0.25^2$, $\mathbb{E}\lbr\sigma_w^2\rbr = 0.1^2$.]{\includegraphics[width=3.2in]{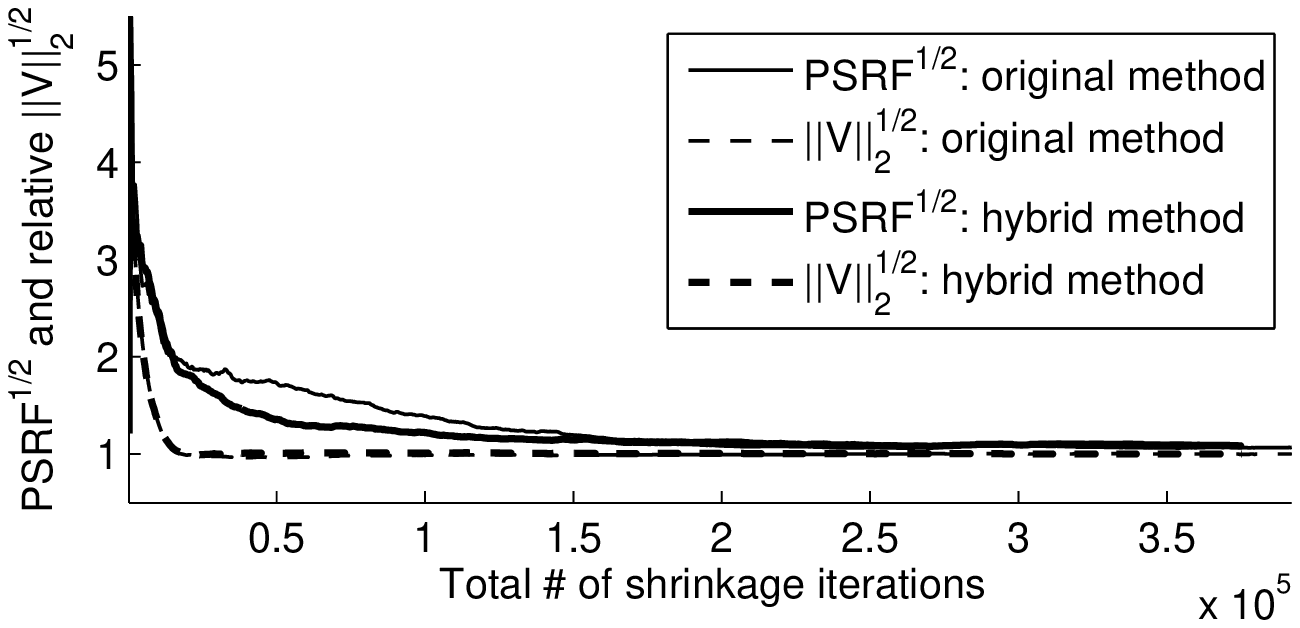}\label{fig:shrinkcomp-2}}
\caption{One-hundred chains of the Gibbs/slice sampler are run for $1000$ iterations each, and the convergence as a function of the total number of shrinkage iterations is measured by the square root of the PSRF and $\|\hat{\mathbf{V}}\|_2^{1/2}$. The values of $\alpha_x$, $\beta_x$, $\alpha_z$, $\beta_z$, $\alpha_w$, and $\beta_w$ are determined for the expected values of $\sigma_x^2$, $\sigma_z^2$, and $\sigma_w^2$ using~\eqref{eq:intro_alphabetax} and~\eqref{eq:intro_alphabetazw}. For a given number of shrinkage iterations, the hybrid rejection-midpoint-threshold method ($\tau = 25$) outperforms the original rejection-based shrinkage method in~\protect\subref{fig:shrinkcomp-1} and performs equally well in~\protect\subref{fig:shrinkcomp-2}.\label{fig:shrinkcomp}}
\end{figure}
To compare these methods in the context of the jitter mitigation, we monitor the convergence of the complete Gibbs sampler, using the shrinkage methods described above. While the combined method obviously shrinks the slice much faster than the original method, the hybrid method's increased speed must offset any increased correlation in the accepted samples in order to be useful. In Figure~\ref{fig:shrinkcomp}, the convergence metrics $\text{PSRF}^{1/2}$ and $\|\hat{\mathbf{V}}\|_2^{1/2}$ are plotted as a function of the total number of shrinkage iterations performed. The convergence rate of the two shrinkage methods are very similar, but in some cases, as shown in Figure~\ref{fig:shrinkcomp-1}, the hybrid method outperforms the original shrinkage method.

To summarize, pseudocode of the slice sampling algorithm using either shrinkage method to generate realizations of $z_n$ is written in Algorithm~\ref{alg:nl_znslicesamp}.
\begin{algorithmDSW}
\begin{algorithmic}
\REQUIRE Previous value $z_n^{(i-1)}$, $\mathbf{x}$, $\sigma_x^2$, $\sigma_z^2$, $\sigma_w^2$, $y_n$, threshold $\tau \geq 0$
\STATE Choose $u \sim U([0,\tilde{p}(z_n^{(i-1)}\mid \mathbf{x},\sigma_x^2,\sigma_z^2,\sigma_w^2,y_n)])$ (see~\eqref{eq:nl_pzncond}).
\STATE Compute initial interval $[L,R]$ according to~\eqref{eq:nl_slicezrange}.
\REPEAT[This is the ``shrinkage'' algorithm from~\cite{Neal03}.]
\STATE Choose $z \sim U([L,R])$.
\IF{$\tilde{p}(z\mid \mathbf{x},y_n,\sigma_x^2,\sigma_z^2,\sigma_w^2) < u$}
\IF{$z < z_n^{(i-1)}$}
\STATE $L \leftarrow z$.
\ELSE
\STATE $R \leftarrow z$.
\ENDIF
\ENDIF
\IF[(Optional) midpoint-threshold modification from rejoinder in~\cite{Neal03}.]{$\tilde{p}(z\mid\mathbf{x},y_n,\sigma_x^2,\sigma_z^2,\sigma_w^2) < e^{-\tau}u$}
\IF{$\frac{1}{2}(L+R) < z_n^{(i-1)}$}
\STATE $L \leftarrow \frac{1}{2}(L+R)$.
\ELSE
\STATE $R \leftarrow \frac{1}{2}(L+R)$.
\ENDIF
\ENDIF
\UNTIL{$\tilde{p}(z\mid \mathbf{x},\sigma_x^2,\sigma_z^2,\sigma_w^2,y_n) \geq u$.}
\RETURN $z$
\end{algorithmic}
\caption{Algorithm for computing $z_n$ with slice sampling.}
\label{alg:nl_znslicesamp}
\end{algorithmDSW}

\subsection{Generating $\mathbf{x}$, $\sigma_x^2$, $\sigma_z^2$, and $\sigma_w^2$}

The full conditional distribution on $x_k$ does depend on the other signal parameters $\mathbf{x}_{\backslash k}$:
\begin{equation}
\begin{split}
p(x_k\mid \mathbf{x}_{\backslash k},\mathbf{z},\sigma_x^2,\sigma_w^2,\sigma_z^2,\mathbf{y}) &= \frac{p(\mathbf{y}\mid \mathbf{z},\mathbf{x},\sigma_w^2)p(\mathbf{z},\sigma_z^2)p(\mathbf{x}\mid\sigma_x^2)p(\sigma_x^2)p(\sigma_w^2)}{p(\mathbf{x}_{\backslash k},\mathbf{z},\sigma_x^2,\sigma_w^2,\sigma_z^2,\mathbf{y})}\\
&\propto \mathcal{N}(\mathbf{y};\mathbf{H(z)x},\sigma_w^2\mathbf{I})\mathcal{N}(x_k;0,\sigma_x^2).\end{split}\label{eq:nl_pxkcond}
\end{equation}
Grouping correlated variables accelerates Gibbs sampler convergence, and the random vector $\mathbf{x}$ can still be generated in one simple step since
\begin{equation}
p(\mathbf{x}\mid\mathbf{z},\sigma_x^2,\sigma_w^2,\sigma_z^2,\mathbf{y}) \propto \mathcal{N}(\mathbf{y};\mathbf{H(z)x},\sigma_w^2\mathbf{I})\mathcal{N}(\mathbf{x};\mathbf{0},\sigma_x^2\mathbf{I})\label{eq:nl_pxcond}
\end{equation}
implies the posterior distribution of $\mathbf{x}$ is just multivariate normal with mean
\begin{align}
\boldsymbol{\mu}_{\mathbf{x}} &= \boldsymbol{\Lambda}_{\mathbf{x}}\frac{\mathbf{H(z)}^T\mathbf{y}}{\sigma_w^2}\label{eq:nl_mux}
\intertext{and covariance matrix}
\boldsymbol{\Lambda}_{\mathbf{x}} &= \sigma_w^2[\mathbf{H(z)}^T\mathbf{H(z)}+\frac{\sigma_w^2}{\sigma_x^2}\mathbf{I}]^{-1}.\label{eq:nl_lambdax}
\end{align}

The Gibbs sampler easily handles the variances $\sigma_x^2$, $\sigma_z^2$, or $\sigma_w^2$ being random variables. The generation of realizations of $z_n$ and $x_k$ proceeds using the previous iteration's estimates of $\sigma_x^2$, $\sigma_z^2$, and $\sigma_w^2$ instead of the true variances. Each cycle of the Gibbs sampler generates realizations of $\sigma_x^2$, $\sigma_z^2$ and $\sigma_w^2$ using the observations $\mathbf{y}$ and the current iteration's values of $\mathbf{z}$ and $\mathbf{x}$. The Gibbs sampler algorithm shown in Algorithm~\ref{alg:nl_gibbs} generates realizations from the posterior pdfs for $\sigma_x^2$, $\sigma_z^2$, and $\sigma_w^2$. Using Bayes rule and the independence of $z_n$ and $w_n$, these conditional pdfs are
\begin{align}
\begin{split}
p(\sigma_x^2\mid \mathbf{x},\mathbf{z},\mathbf{y},\sigma_z^2,\sigma_w^2) &= p(\sigma_x^2\mid \mathbf{x}) = \frac{p(\mathbf{x}\mid \sigma_x^2)p(\sigma_x^2)}{p(\mathbf{x})}\\
&\propto \mathcal{N}(\mathbf{x};\mathbf{0},\sigma_x^2\mathbf{I})\mathcal{IG}(\sigma_x^2;\alpha_x,\beta_x),\end{split}\label{eq:nl_psigmaxcond}\\
\begin{split}
p(\sigma_z^2\mid \mathbf{x},\mathbf{z},\mathbf{y},\sigma_x^2,\sigma_w^2) &= p(\sigma_z^2\mid \mathbf{z}) = \frac{p(\mathbf{z}\mid \sigma_z^2)p(\sigma_z^2)}{p(\mathbf{z})}\\
&\propto \mathcal{N}(\mathbf{z};\mathbf{0},\sigma_z^2\mathbf{I})\mathcal{IG}(\sigma_z^2;\alpha_z,\beta_z),\end{split}\label{eq:nl_psigmazcond}\\
\intertext{and}
\begin{split}
p(\sigma_w^2\mid \mathbf{x},\mathbf{z},\mathbf{y},\sigma_x^2,\sigma_z^2) &= p(\sigma_w^2\mid \mathbf{x},\mathbf{z},\mathbf{y}) = \frac{p(\mathbf{y}\mid\mathbf{x},\mathbf{z},\sigma_w^2)p(\mathbf{x})p(\mathbf{z})p(\sigma_w^2)}{p(\mathbf{y},\mathbf{x},\mathbf{z})}\\
&\propto \mathcal{N}(\mathbf{y};\mathbf{H(z)x},\sigma_w^2\mathbf{I})\mathcal{IG}(\sigma_w^2;\alpha_w,\beta_w).\end{split}\label{eq:nl_psigmawcond}
\end{align}
The inverse Gamma distribution is the conjugate prior for the variance parameter of a Normal distribution (see~\cite{Gelman04}). Therefore, the posterior distribution is also an inverse Gamma distribution. Specifically, $p(\sigma_x^2\mid\mathbf{x},\mathbf{z},\mathbf{y},\sigma_z^2,\sigma_w^2) = \mathcal{IG}(\sigma_x^2;\alpha_x',\beta_x')$, where
\begin{align}
\alpha_x' &= \alpha_x + \frac{K}{2};&\quad\beta_x' &= \beta_x + \frac{\|\mathbf{x}\|_2^2}{2}.\label{eq:nl_postalphabetax}
\end{align}
Similarly the hyperparameters for the posterior inverse Gamma distributions on $\sigma_z^2$ and $\sigma_w^2$ are
\begin{align}
\alpha_z' &= \alpha_z + \frac{N}{2};&\quad\beta_z' &= \beta_z + \frac{\|\mathbf{z}\|_2^2}{2};\label{eq:nl_postalphabetaz}\\
\alpha_w' &= \alpha_w + \frac{N}{2};&\quad\beta_w' &= \beta_w + \frac{\|\mathbf{y-H(z)x}\|_2^2}{2}.\label{eq:nl_postalphabetaw}
\end{align}
Thus, generating realizations of $\sigma_z^2$ or $\sigma_w^2$ using such a prior is as simple as taking the inverse of realizations of a gamma distribution with the proper choice of hyperparameters. For those who prefer a non-informative prior, the Jeffreys priors for $\sigma_x^2$, $\sigma_z^2$, and $\sigma_w^2$ are $p(\sigma_x^2) = 1/\sigma_x^2$, $p(\sigma_z^2) = 1/\sigma_z^2$, and $p(\sigma_w^2) = 1/\sigma_w^2$. Although these priors are improper distributions, they are equivalent to inverse Gamma distributions with $\alpha = \beta = 0$, so the associated posterior distributions are proper inverse Gamma distributed with the parameters described above.

Once enough samples have been taken so that the current state of the Markov chain is sufficiently close to the steady state, the Gibbs sampling theory tells us that further samples drawn from the chain can be treated as if they were drawn from the joint posterior distribution directly. Thus, these additional samples can be averaged to approximate the Bayes MMSE estimator. In the complete Gibbs sampler in Algorithm~\ref{alg:nl_gibbs}, $I_b$ represents the ``burn-in time,'' the number of iterations until the Markov chain has approximately reached its steady state, and $I$ represents the number of samples to generate after convergence, which are averaged to form the MMSE estimates.
\begin{algorithmDSW}
\begin{algorithmic}
\REQUIRE $\mathbf{y}, I, I_b$
\STATE $\mathbf{z}^{(0)} \leftarrow \mathbf{0}$; $\mathbf{x}^{(0)} \leftarrow \mathbf{\hat{x}}_{\text{LMMSE}\mid\mathbf{z}=\mathbf{0}}(\mathbf{y})$ from~\eqref{eq:lin_lmmseestnoz}; ${\sigma_x^2}^{(0)} \leftarrow 1$; ${\sigma_z^2}^{(0)} \leftarrow 0.01$; ${\sigma_w^2}^{(0)} \leftarrow 0.01$
\FOR{$i = 1:I+I_b$}
\FOR{$n = 0:N-1$}
\STATE Generate $z_n^{(i)}$ using slice sampling in Algorithm~\ref{alg:nl_znslicesamp}.
\ENDFOR
\STATE Generate $\mathbf{x}^{(i)}$ from $\mathcal{N}(\boldsymbol{\mu}_{\mathbf{x}},\boldsymbol{\Lambda}_{\mathbf{x}})$ using~\eqref{eq:nl_mux} and~\eqref{eq:nl_lambdax}.
\STATE Generate ${\sigma_x^2}^{(i)}$ from $\mathcal{IG}(\alpha_x',\beta_x')$ using~\eqref{eq:nl_postalphabetax}.\\
\STATE Generate ${\sigma_z^2}^{(i)}$ from $\mathcal{IG}(\alpha_z',\beta_z')$ using~\eqref{eq:nl_postalphabetaz}.\\
\STATE Generate ${\sigma_w^2}^{(i)}$ from $\mathcal{IG}(\alpha_w',\beta_w')$ using~\eqref{eq:nl_postalphabetaw}.\\
\ENDFOR
\STATE $\mathbf{\hat{x}} \leftarrow \frac{1}{I}\sum_{i=I_b+1}^{I_b+I} \mathbf{x}^{(i)}$
\STATE $\mathbf{\hat{z}} \leftarrow \frac{1}{I}\sum_{i=I_b+1}^{I_b+I} \mathbf{z}^{(i)}$
\STATE $\hat{\sigma}_x^2 \leftarrow \frac{1}{I}\sum_{i=I_b+1}^{I_b+I} {\sigma_x^2}^{(i)}$
\STATE $\hat{\sigma}_z^2 \leftarrow \frac{1}{I}\sum_{i=I_b+1}^{I_b+I} {\sigma_z^2}^{(i)}$
\STATE $\hat{\sigma}_w^2 \leftarrow \frac{1}{I}\sum_{i=I_b+1}^{I_b+I} {\sigma_w^2}^{(i)}$
\RETURN $\mathbf{\hat{x}}$, $\mathbf{\hat{z}}$, $\hat{\sigma}_x^2$, $\hat{\sigma}_z^2$, $\hat{\sigma}_w^2$
\end{algorithmic}
\caption{Pseudocode for the Gibbs sampler modified to use slice sampling for the $z_n$'s.}
\label{alg:nl_gibbs}
\end{algorithmDSW}

\section{Simulation Results}\label{sec:simresults}

In this section, both the convergence behavior and the performance of the Gibbs/slice sampler are analyzed. Using Matlab, a $K$-parameter signal and $N = K\,M$ samples of that signal are generated with pseudo-random jitter and additive noise; $M$ is the oversampling factor. Then, implementations of the Gibbs/slice sampler, as well as the linear MMSE estimator in~\eqref{eq:lin_lmmseest}, the no-jitter linear estimator in~\eqref{eq:lin_lmmseestnoz}, and the EM algorithm developed in~\cite{WellerClassical} for approximating the ML estimator are applied to the samples. The adaptation of the EM algorithm to random $\sigma_w^2$ and $\sigma_z^2$ is described in the appendix; however, the EM algorithm with known $\sigma_w^2$ and $\sigma_z^2$ is used in these simulations because adapting to random variances dramatically increases the computational cost, and the difference in MSE is negligible. These algorithms are studied in detail for periodic bandlimited signals with uniformly distributed signal parameters in~\cite{WellerThesis}, and in this work, a similar analysis is performed to analyze the convergence and sensitivity to initial conditions of the proposed algorithms. This analysis is also similar to that performed in~\cite{WellerClassical} for the EM algorithm approximation to the ML estimator of the non-Bayesian version of this paper's problem formulation.

\subsection{Convergence Analysis}

As a Markov chain Monte Carlo method, the Gibbs/slice sampler converges to the appropriate posterior distribution under certain conditions (see~\cite{Robert04}); as long as the sequence generated by sampling from the steady-state distribution $p(\mathbf{x,z},\sigma_x^2,\sigma_z^2,\sigma_w^2\mid\mathbf{y})$ is ergodic, the samples can be averaged to approximate the Bayes MMSE estimate of the signal parameters. In addition, the steady-state distribution of an irreducible chain is unique, so the choice of initialization should not impact the final estimate generated from the steady-state samples. Of course, since the chain only converges to the steady-state in the limit, small transient effects from the initial conditions are evaluated.

\begin{figure}[!t]
\centering
\subfloat[][$K = 10$, $\mathbb{E}\lbr\sigma_z^2\rbr = 0.25^2$, $\mathbb{E}\lbr\sigma_w^2\rbr = 0.1^2$, $M$ varies.]{\includegraphics[width=3.2in]{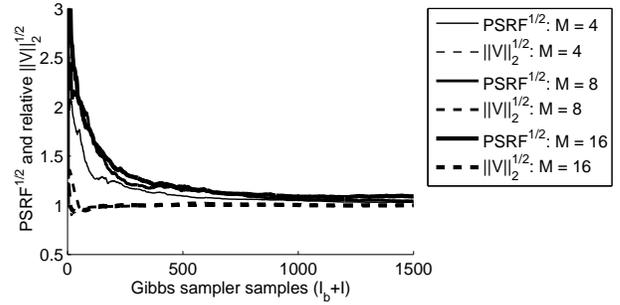}\label{fig:convergePSRF-1}}\\
\subfloat[][$K = 10$, $M = 4$, $\mathbb{E}\lbr\sigma_w^2\rbr = 0.1^2$, $\mathbb{E}\lbr\sigma_z^2\rbr$ varies.]{\includegraphics[width=3.2in]{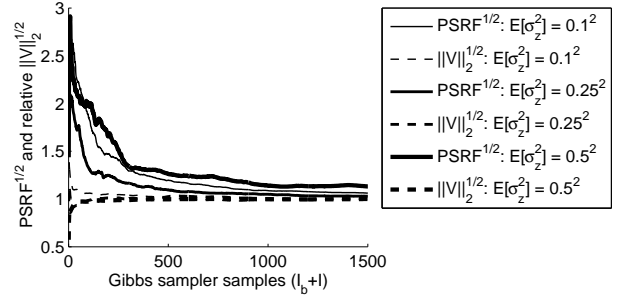}\label{fig:convergePSRF-2}}\\
\subfloat[][$K = 10$, $M = 4$, $\mathbb{E}\lbr\sigma_z^2\rbr = 0.25^2$, $\mathbb{E}\lbr\sigma_w^2\rbr$ varies.]{\includegraphics[width=3.2in]{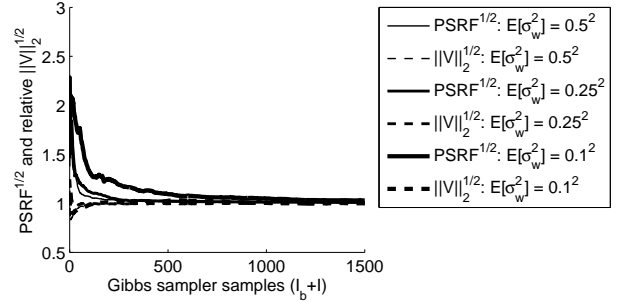}\label{fig:convergePSRF-3}}
\caption{The convergence of the Gibbs/slice sampler ($100$ chains, $1500$ samples) as a function of the number of samples $I_b+I$ is measured by the $\text{PSRF}^{1/2}$ and $\|\mathbf{\hat{V}}\|_2^{1/2}$ convergence metrics. The $\|\mathbf{\hat{V}}\|_2^{1/2}$ values are normalized by the final value for each curve. The parameters $\alpha_x$, $\beta_x$, $\alpha_z$, $\beta_z$, $\alpha_w$, and $\beta_w$ are determined using~\eqref{eq:intro_alphabetax} and~\eqref{eq:intro_alphabetazw}. The rate of convergence depends on the choice of parameters, as demonstrated in the above plots.\label{fig:convergePSRF}}
\end{figure}
The rate of convergence of the Gibbs/slice sampler, as measured by the $\|\mathbf{\hat{V}}\|_2^{1/2}$ and the square root of the PSRF, is shown in Figure~\ref{fig:convergePSRF}. The results suggest that increasing the oversampling factor $M$ or the jitter variance $\sigma_z^2$ or decreasing the additive noise variance $\sigma_w^2$ slows the rate of convergence. In most cases, the Markov chain appears to reach a steady state within $500$ iterations; thus, we set $I_b = 500$ iterations (see Algorithm~\ref{alg:nl_gibbs}) for the tests that follow.

\begin{figure}[!t]
\centering
\subfloat[][$K = 10$, $\mathbb{E}\lbr\sigma_z^2\rbr = 0.25^2$, $\mathbb{E}\lbr\sigma_w^2\rbr = 0.1^2$, $M$ varies.]{\includegraphics[width=3.2in]{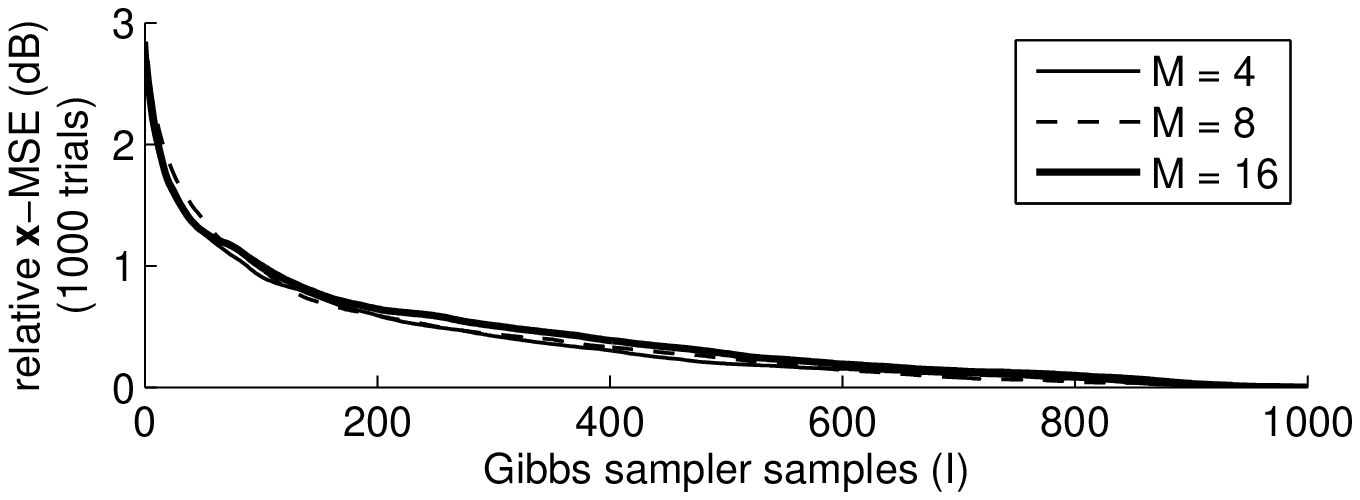}\label{fig:convergeMSE-1}}\\
\subfloat[][$K = 10$, $M = 4$, $\mathbb{E}\lbr\sigma_w^2\rbr = 0.1^2$, $\mathbb{E}\lbr\sigma_z^2\rbr$ varies.]{\includegraphics[width=3.2in]{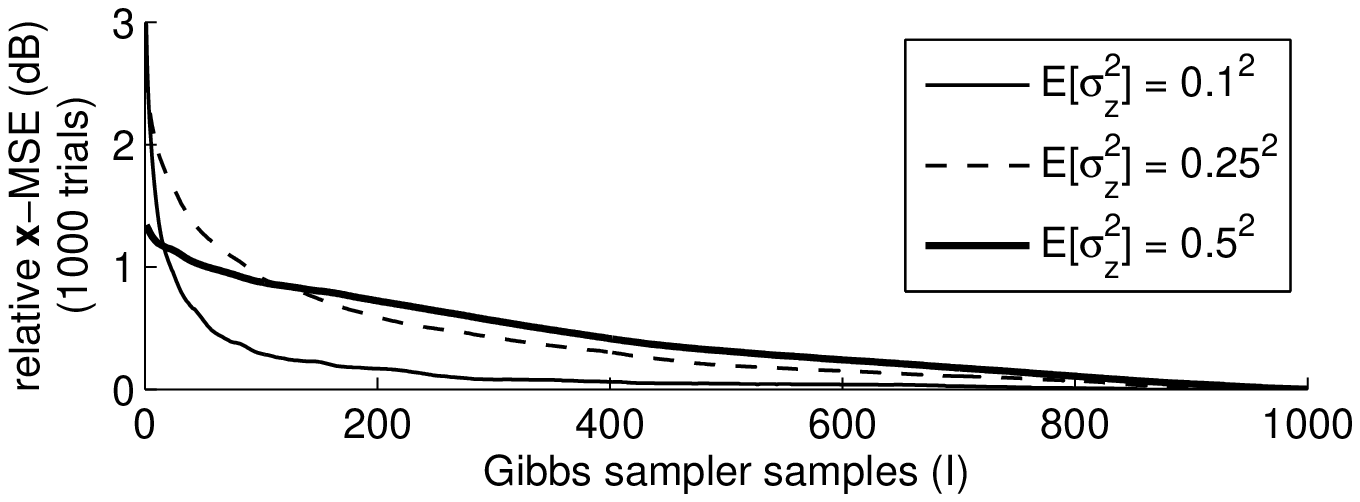}\label{fig:convergeMSE-2}}\\
\subfloat[][$K = 10$, $M = 4$, $\mathbb{E}\lbr\sigma_z^2\rbr = 0.25^2$, $\mathbb{E}\lbr\sigma_w^2\rbr$ varies.]{\includegraphics[width=3.2in]{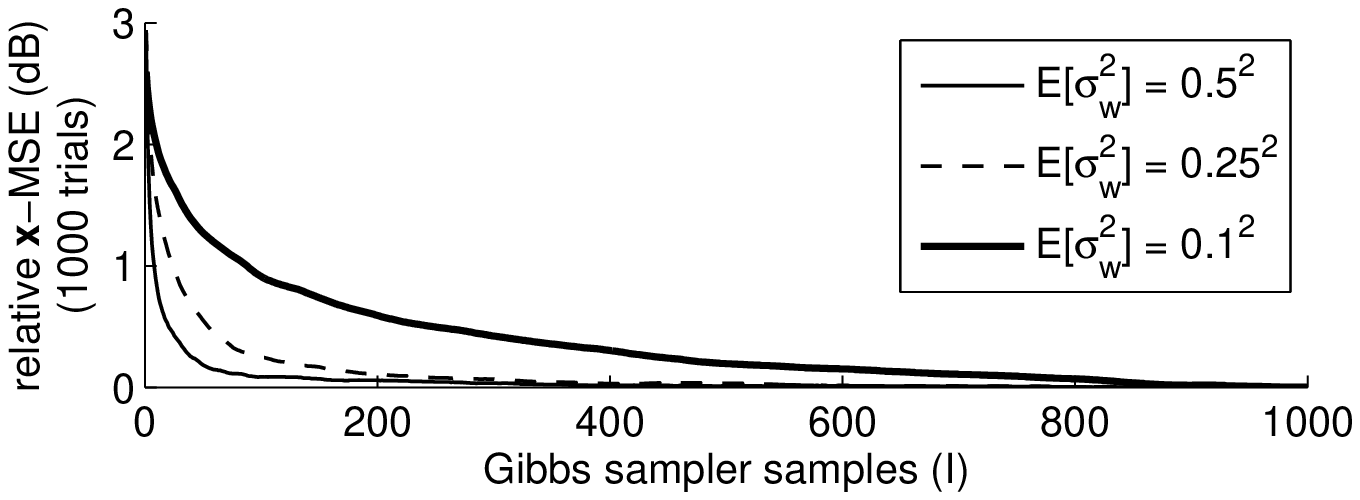}\label{fig:convergeMSE-3}}
\caption{The convergence of the estimator for $\mathbf{x}$ from the Gibbs/slice sampler ($I_b = 500$, $1 \leq I \leq 1000$ samples) is measured from $1000$ trials by the MSE of the Gibbs sampler estimate of $\mathbf{x}$; the MSE is normalized so the MSE for $I = 1000$ samples is $0$ dB. The parameters $\alpha_x$, $\beta_x$, $\alpha_z$, $\beta_z$, $\alpha_w$, and $\beta_w$ are determined using~\eqref{eq:intro_alphabetax} and~\eqref{eq:intro_alphabetazw}. The rate of convergence (when the error line stabilizes) depends on the choice of parameters, as demonstrated in the above plots.\label{fig:convergeMSE}}
\end{figure}
To establish the number of iterations $I$ needed after burn-in, we observe the squared error $\|\mathbf{\hat{x}}_I-\mathbf{x}^*\|_2^2$, where $\mathbf{\hat{x}}_I$ is the $I$th estimate of $\mathbf{x}$, as a function of $I$, for $I$ up to $1000$, and $\mathbf{x}^*$ is the true value of $\mathbf{x}$. Examining the plots in Figure~\ref{fig:convergeMSE}, approximately $500$ iterations are sufficient to achieve a squared error within $0.5$ dB of the asymptotic MSE (as measured by $I = 1000$) for all cases.

\begin{figure}[!t]
\centering
\subfloat[][$K = 10$, $\mathbb{E}\lbr\sigma_z^2\rbr = 0.25^2$, $\mathbb{E}\lbr\sigma_w^2\rbr = 0.25^2$, $M$ varies.]{\includegraphics[width=3.4in]{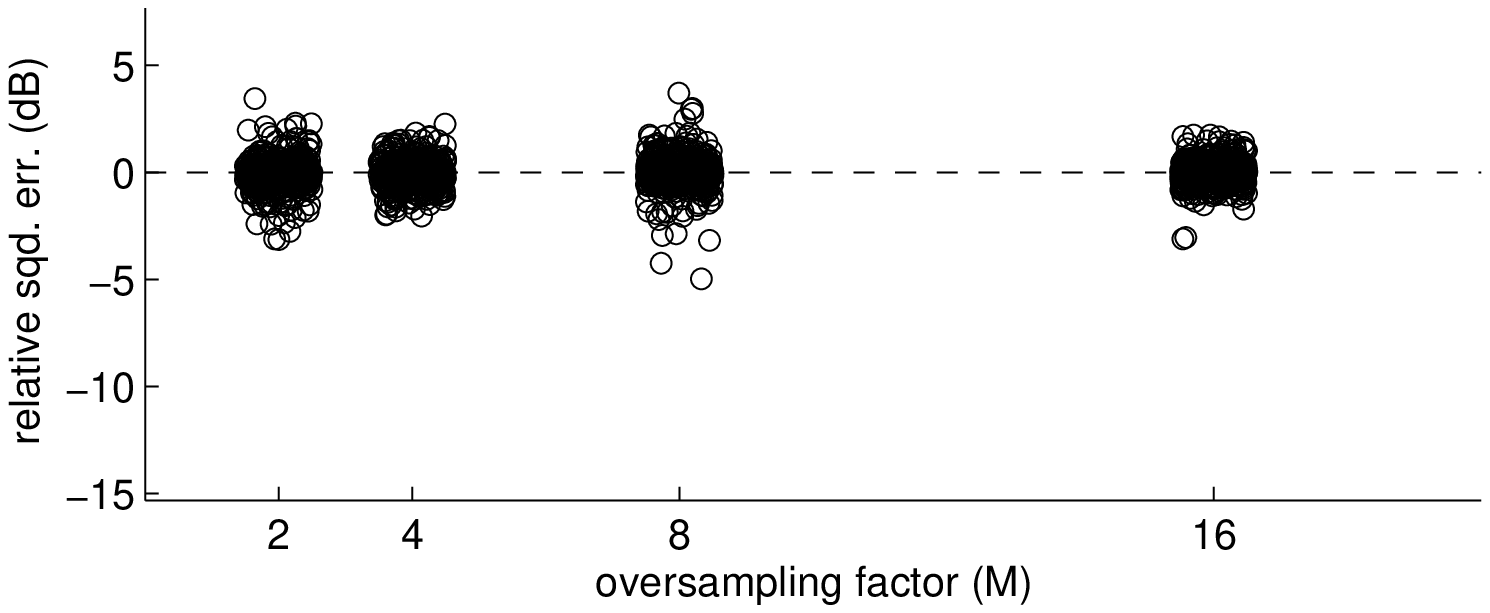}\label{fig:initconds-Ms}}\\
\subfloat[][$K = 10$, $M = 8$, $\mathbb{E}\lbr\sigma_w^2\rbr = 0.25^2$, $\mathbb{E}\lbr\sigma_z^2\rbr$ varies.]{\includegraphics[width=3.4in]{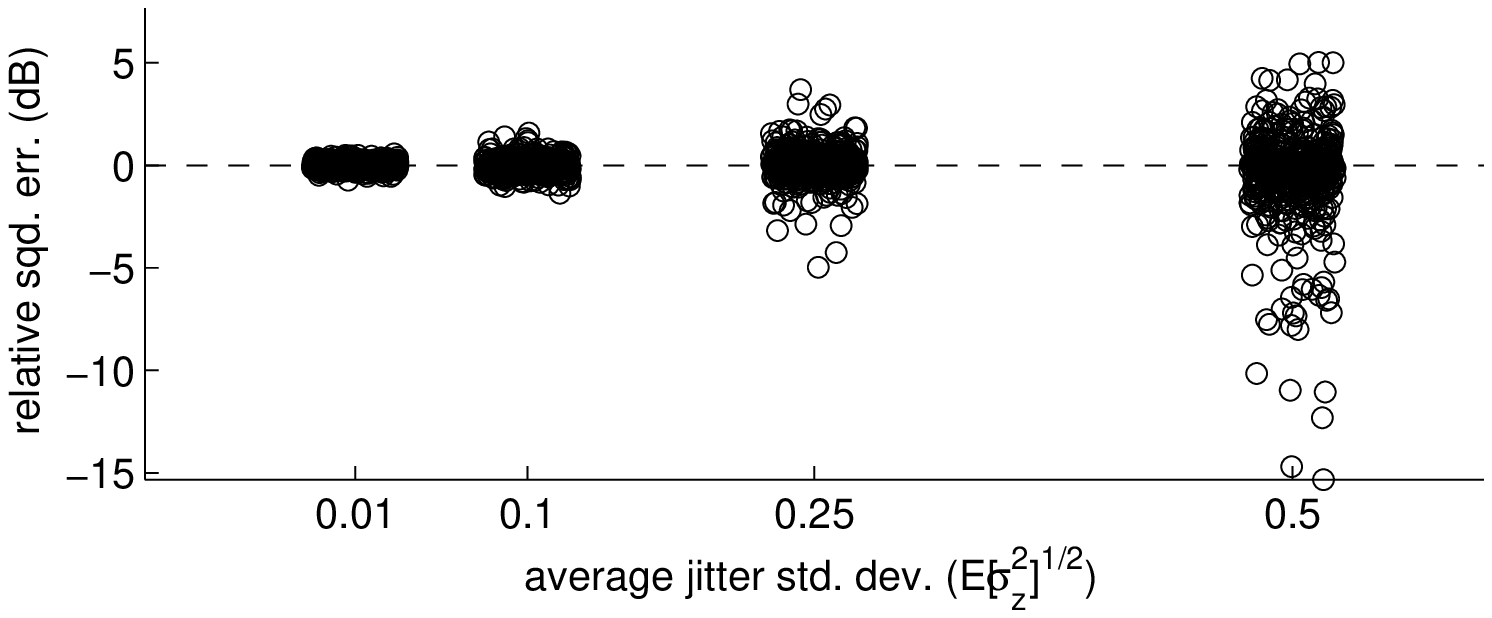}\label{fig:initconds-sigmazs}}\\
\subfloat[][$K = 10$, $M = 8$, $\mathbb{E}\lbr\sigma_z^2\rbr = 0.25^2$, $\mathbb{E}\lbr\sigma_w^2\rbr$ varies.]{\includegraphics[width=3.4in]{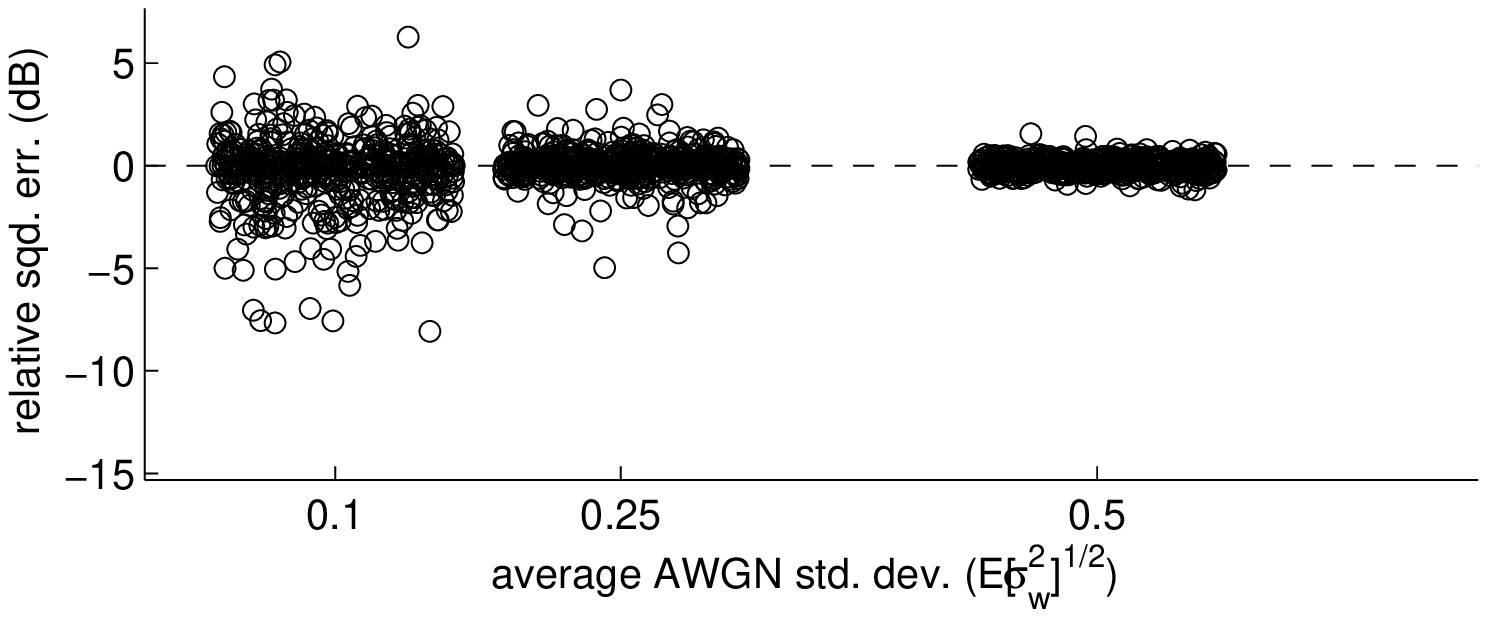}\label{fig:initconds-sigmaws}}
\caption{The effects of varying initial conditions of the Gibbs/slice sampler as a function of oversampling factor~\protect\subref{fig:initconds-Ms}, jitter variance~\protect\subref{fig:initconds-sigmazs}, and additive noise variance~\protect\subref{fig:initconds-sigmaws} are studied by computing the squared errors of the results, for multiple initial conditions, across $50$ trials. The squared errors of the results are normalized relative to the result for initialization with the zero-jitter LMMSE in~\eqref{eq:lin_lmmseestnoz}, so that the squared error of the result for initialization with this linear estimator is $0$ dB. The parameters $\alpha_x$, $\beta_x$, $\alpha_z$, $\beta_z$, $\alpha_w$, and $\beta_w$ are determined using~\eqref{eq:intro_alphabetax} and~\eqref{eq:intro_alphabetazw}.\label{fig:initconds}}
\end{figure}
The sensitivity to initial conditions of the Gibbs/slice sampler is shown in Figure~\ref{fig:initconds} for $I_b = I = 500$. For $50$ trials, the squared error of the Bayes MMSE estimates are measured for ten different choices of initial conditions. The ten choices of initial conditions used are (1) $\sigma_x^{(0)} = 1$, $\sigma_z^{(0)} = \sigma_w^{(0)} = 0.1$, and all $\mathbf{x}^{(0)}$ and $\mathbf{z}^{(0)}$ equal to zero, (2) $\sigma_x^{(0)} = 1$, $\sigma_z^{(0)} = \sigma_w^{(0)} = 0.1$, $\mathbf{z}^{(0)}$ equal to zero, and the no-jitter LMMSE estimate for $\mathbf{x}^{(0)}$, (3) the true values of $\sigma_x^2$, $\sigma_z^2$, $\sigma_w^2$, $\mathbf{z}$, and $\mathbf{x}$, and (4-10) seven choices of random values of $\sigma_x^2$, $\sigma_z^2$, $\sigma_w^2$, $\mathbf{z}$ and the corresponding fixed-jitter LMMSE estimates for $\mathbf{x}$. The squared errors displayed are normalized so that the squared error for the no-jitter LMMSE estimate starting point equals one. Although the Gibbs/slice sampler becomes more sensitive to initial conditions as $\sigma_z$ increases, in all cases, the squared errors for the majority of initial conditions are close to one. Thus, even though the algorithms are still sensitive to initial conditions after the burn-in period, especially for larger jitter variance, the choice of no-jitter LMMSE estimate is about average.

\subsection{Performance Comparisons}

\begin{figure}[!t]
\centering
\subfloat[][$K = 10$, $M = 4$, $\mathbb{E}\lbr\sigma_w^2\rbr = 0.05^2$, $0.01 \leq \mathbb{E}\lbr\sigma_z^2\rbr^{1/2} \leq 0.5$.]{\includegraphics[width=3.45in]{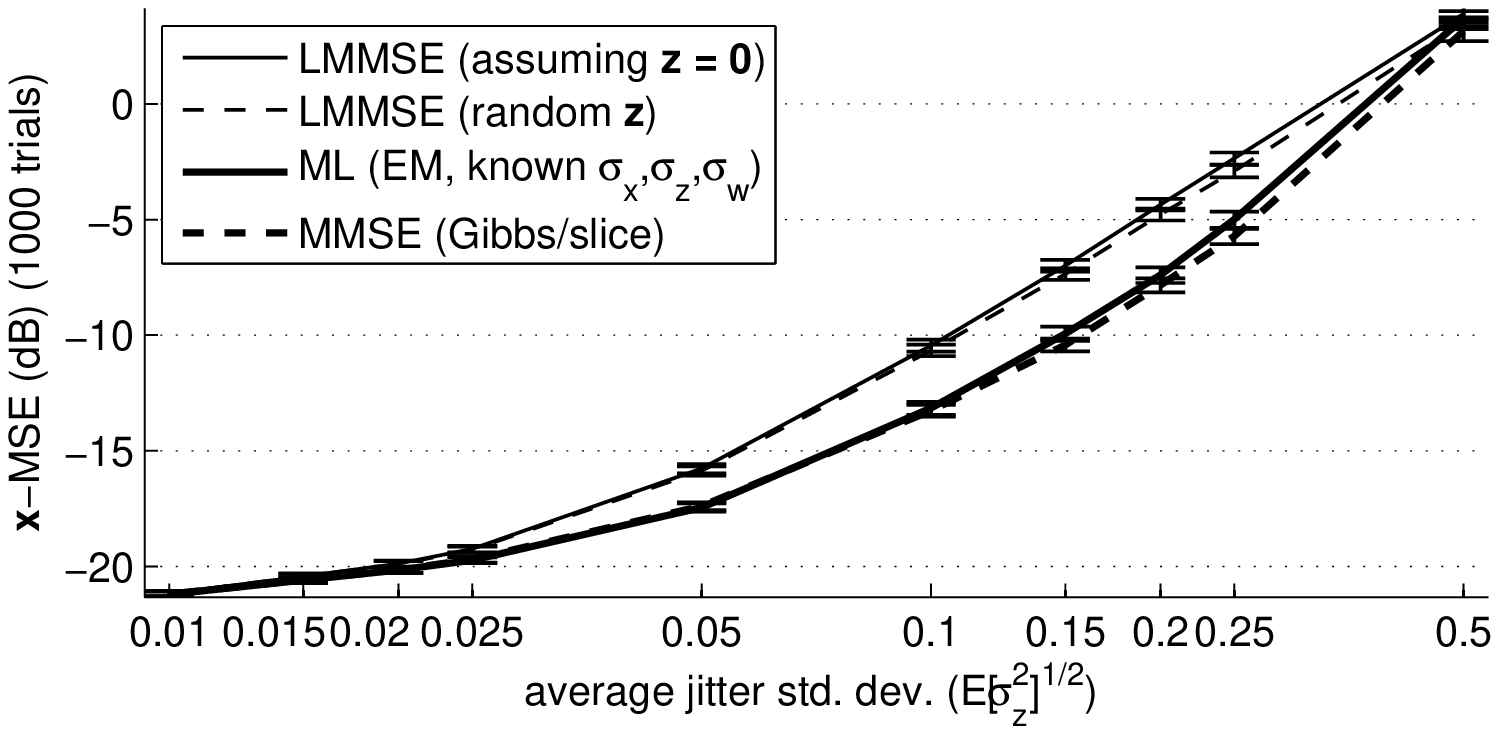}\label{fig:perf-1}}\\
\subfloat[][$K = 10$, $M = 16$, $\mathbb{E}\lbr\sigma_w^2\rbr = 0.05^2$, $0.01 \leq \mathbb{E}\lbr\sigma_z^2\rbr^{1/2} \leq 0.5$.]{\includegraphics[width=3.45in]{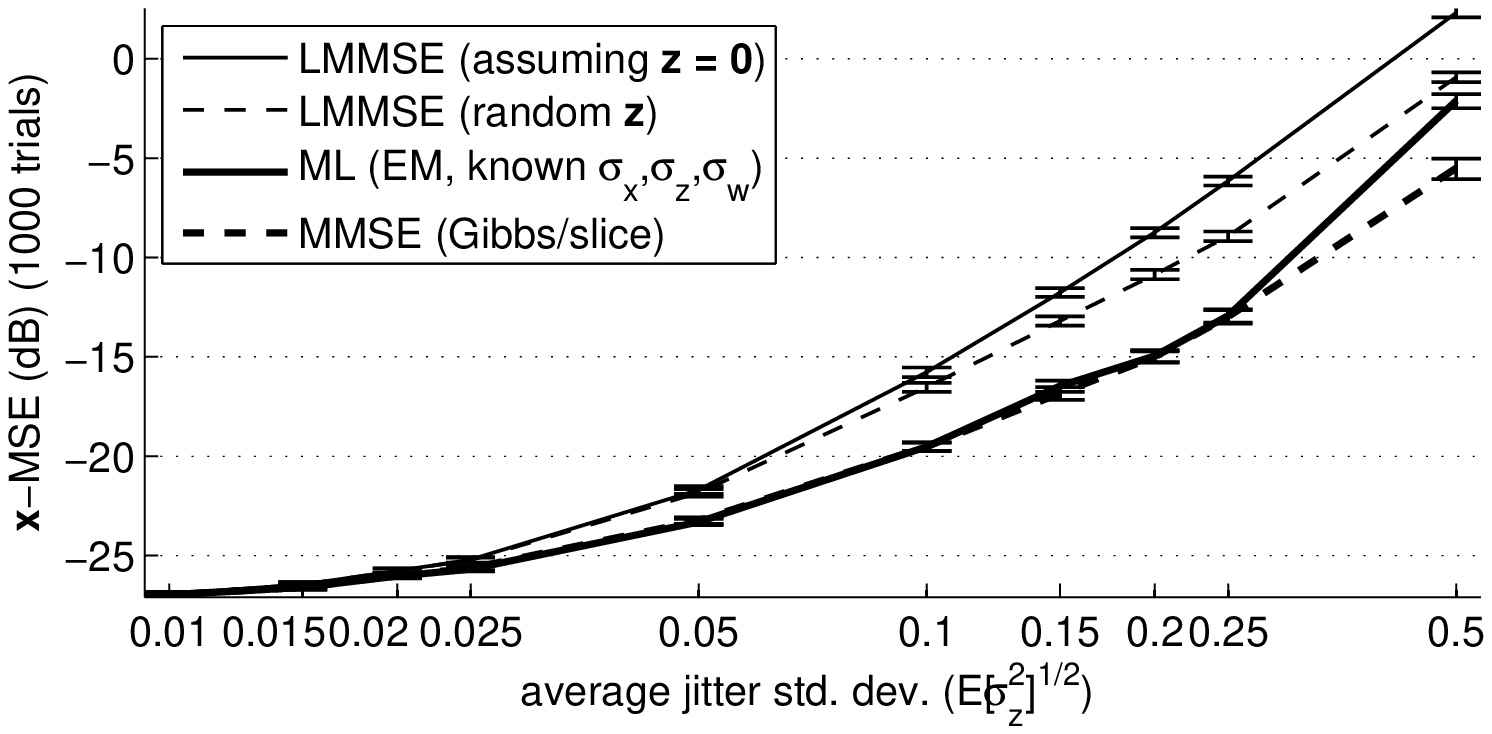}\label{fig:perf-2}}\\
\subfloat[][$K = 10$, $M = 16$, $\mathbb{E}\lbr\sigma_w^2\rbr = 0.025^2$, $0.01 \leq \mathbb{E}\lbr\sigma_z^2\rbr^{1/2} \leq 0.5$.]{\includegraphics[width=3.45in]{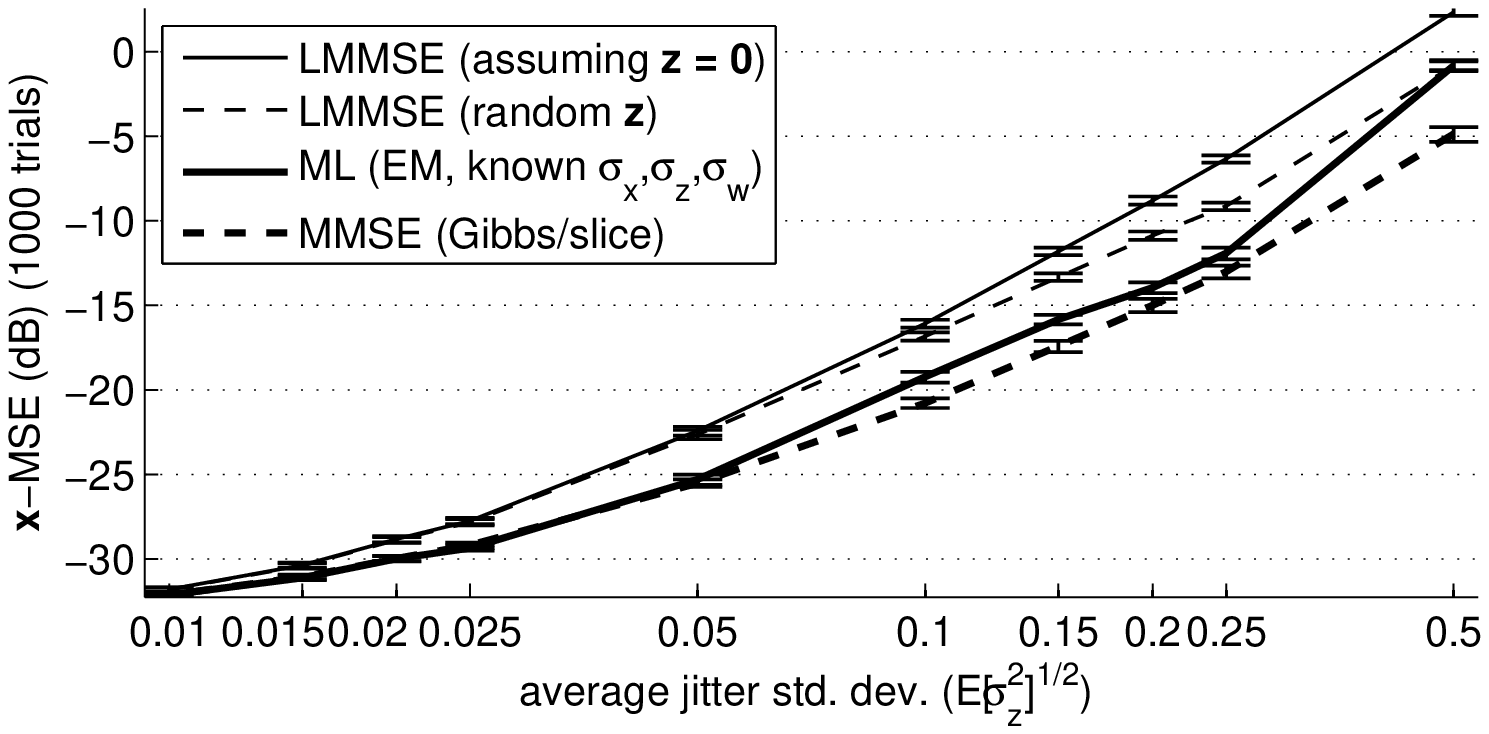}\label{fig:perf-3}}
\caption{The MSE performance of the Bayes MMSE estimator as computed using the Gibbs/slice sampler is compared against both the unbiased linear MMSE estimator~\eqref{eq:lin_lmmseest} and the no-jitter linear MMSE estimator~\eqref{eq:lin_lmmseestnoz}, as well as the EM algorithm approximation to the ML estimator from~\cite{WellerClassical}. The values of $\alpha_x$, $\beta_x$, $\alpha_z$, $\beta_z$, $\alpha_w$, and $\beta_w$ are determined for the average $\sigma_x^2$, $\sigma_z^2$, and $\sigma_w^2$ using~\eqref{eq:intro_alphabetax} and~\eqref{eq:intro_alphabetazw}. The EM algorithm uses the true values of $\sigma_x^2$, $\sigma_z^2$, and $\sigma_w^2$, while the linear estimators and the Gibbs/slice sampler treat $\sigma_x^2$, $\sigma_z^2$, and $\sigma_w^2$ as random variables. The error bars above and below each data point for the estimators delineate the $95\%$ confidence intervals for those data points.
\label{fig:perf}}
\end{figure}
In Figure~\ref{fig:perf}, the performance of the Gibbs/slice sampler is compared against the linear MMSE and no-jitter linear MMSE estimators and the EM algorithm approximation to the ML estimator derived in~\cite{WellerClassical}. The MSE performances are plotted for different values of $M$, $\sigma_z$, and $\sigma_w$ to demonstrate the effect of increasing $M$, increasing $\sigma_z$, or decreasing $\sigma_w$ on the relative MSE performances. Comparing the Gibbs/slice sampler Bayes MMSE estimate against the linear estimator, the Gibbs/slice sampler outperforms the linear MMSE estimator for a large range of $\sigma_z$, a difference that becomes more pronounced with higher oversampling $M$. In addition, the results suggest that the Gibbs/slice sampler outperforms classical estimation, especially for higher jitter variances.

We also compare computation times for the EM algorithm and the Gibbs/slice sampler. Both converge more slowly for higher jitter and lower additive noise, and greater oversampling also lengthens computation. In the case of $K = 10$, $M = 16$, $\mathbb{E}[\sigma_z^2] = 0.5^2$, and $\mathbb{E}[\sigma_w^2] = 0.025^2$, the EM algorithm with known $\sigma_z^2$ and $\sigma_w^2$ requires $1.6$ seconds per trial on average, the EM algorithm for random noise variances requires $24$ seconds, and the Gibbs/slice sampler requires $3.1$ seconds on average. In only an eighth the time, the Gibbs/slice sampler achieves greater MSE performance than the EM algorithm.

\begin{figure}[!t]
\centering
\subfloat[][$K = 10$, $M$ varies, $\mathbb{E}\lbr\sigma_w^2\rbr = 0.025^2$, $\frac{1}{2}\mathbb{E}\lbr\sigma_w^2\rbr^{1/2} \leq \mathbb{E}\lbr\sigma_z^2\rbr^{1/2} \leq 0.5$.]{\includegraphics[width=3.45in]{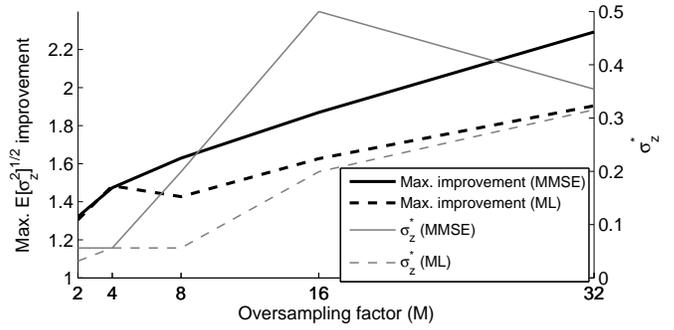}\label{fig:jitterimprove-Ms}}\\
\subfloat[][$K = 10$, $M = 16$, $\mathbb{E}\lbr\sigma_w^2\rbr^{1/2}$ varies, $\frac{1}{2}\mathbb{E}\lbr\sigma_w^2\rbr^{1/2} \leq \mathbb{E}\lbr\sigma_z^2\rbr^{1/2} \leq 0.5$.]{\includegraphics[width=3.45in]{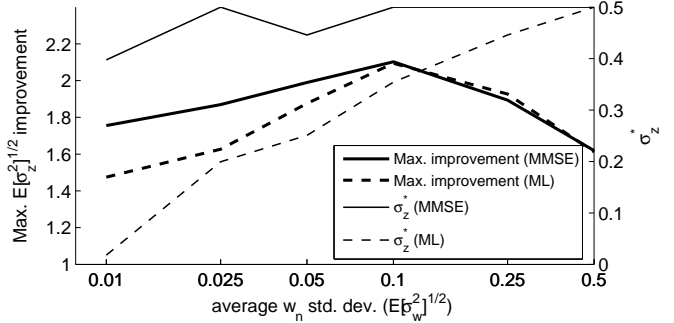}\label{fig:jitterimprove-sigmaws}}
\caption{Jitter improvement from using MMSE (Gibbs/slice sampler) and ML estimators (EM algorithm with known $\sigma_x$, $\sigma_z$, and $\sigma_w$) is measured by interpolating the maximum factor of improvement in jitter tolerance, measured by $\mathbb{E}\left[\sigma_z^2\right]^{1/2}$, relative to using no-jitter LMMSE reconstruction. Holding $\mathbb{E}\left[\sigma_w^2\right]$ fixed, \protect\subref{fig:jitterimprove-Ms} shows the trend in maximum improvement as $M$ increases, and \protect\subref{fig:jitterimprove-sigmaws} shows the trend in maximum improvement as $\mathbb{E}\left[\sigma_w^2\right]^{1/2}$ increases while holding $M$ fixed. The jitter standard deviation $\sigma_z^*$ corresponding to this maximum improvement for the MMSE and ML estimators is plotted on the same axes.\label{fig:jitterimproves}}
\end{figure}
To understand the effectiveness of these methods in mitigating jitter, the difference in jitter variance as a function of target MSE is computed based on the performance results and the maximum observed differences (for $\mathbb{E}[\sigma_z^2]^{1/2} \geq \frac{1}{2}\mathbb{E}[\sigma_w^2]^{1/2}$, to avoid the region where the MSE plots are flat) are compared for different values of $M$ and $\mathbb{E}[\sigma_w^2]^{1/2}$. The resulting trends portrayed in Figure~\ref{fig:jitterimproves} demonstrate that greater improvement is achievable with increased oversampling $M$, and small additive noise variance $\mathbb{E}[\sigma_w^2]$. In addition, the Gibbs/slice sampler outperforms the classical ML estimator (as approximated by the EM algorithm in~\cite{WellerClassical}) at high jitter, increasing the factor of improvement, especially in the high oversampling and low additive noise variance regimes.

\section{Conclusion}\label{sec:conclusion}

The results displayed in this paper suggest that post-processing jittered samples with a nonlinear algorithm like Gibbs/slice sampling mitigates the effect of sampling jitter on the total sampling error. In particular, the expected jitter standard deviation can be increased by as much as a factor of $2.2$, enabling substantial power savings in the analog circuitry when compared against linear post-processing or classical nonlinear post-processing (the EM algorithm). Such power savings may enable significant improvements in battery life for implantable cardiac pacemakers and enable the inclusion of ADCs in ultra-low power devices.

Like the EM algorithm proposed in~\cite{WellerClassical}, the Gibbs/slice sampler proposed here suffers from relatively high computational complexity and an iterative nature, which may be unsuitable for embedded applications. Developments in polynomial estimators, such as the Volterra filter-like polynomial estimators described in~\cite{WellerICASSP}, may yield similar performance to the Gibbs/slice sampler proposed here, at least for low levels of oversampling, without such high online computational cost. Further investigation is warranted in developing these and similar approaches for post-processing jittered samples in ADCs. Nevertheless, for off-chip post-processing of jittered samples, the nonlinear Bayesian Gibbs/slice sampler presented here outperforms both linear MMSE estimator and the nonlinear classical EM algorithm approximation to the ML estimator.

\appendix[ML Estimation with Random Variances]

In~\cite{WellerClassical}, the EM algorithm approximation to the ML estimator is derived in the classical setting for known variances $\sigma_z^2$ and $\sigma_w^2$. To adapt the method for random variances, we introduce $\sigma_z^2$ and $\sigma_w^2$ as latent variables:
\begin{equation}
\mathbf{\hat{x}}^{(i)} = \argmax_{\mathbf{x}} \mathbb{E}\left[\log p(\mathbf{y,z},\sigma_z^2,\sigma_w^2;\mathbf{x})\mid\mathbf{y};\mathbf{\hat{x}}^{(i-1)}\right].\label{eq:ml_emalg}
\end{equation}
By conditional independence,
\begin{equation}
\begin{split}
p(\mathbf{y,z},\sigma_z^2,\sigma_w^2;\mathbf{x}) &= p(\mathbf{y}\mid\mathbf{z},\sigma_w^2;\mathbf{x})p(\mathbf{z}\mid\sigma_z^2)p(\sigma_z^2)p(\sigma_w^2)\\
&= \mathcal{N}(\mathbf{y};\mathbf{H(z)x},\sigma_w^2\mathbf{I})p(\mathbf{z}\mid\sigma_z^2)p(\sigma_z^2)p(\sigma_w^2).\end{split}
\end{equation}
The terms not involving $\mathbf{x}$ are unnecessary, since we are differentiating with respect to $\mathbf{x}$ in the next step. The derivative of the expectation in~\eqref{eq:ml_emalg} is
\begin{equation}
\mathbb{E}\left[-\frac{2\mathbf{H}^T\mathbf{(z)}(\mathbf{H(z)x-y})}{2\sigma_w^2}\mid\mathbf{y};\mathbf{\hat{x}}^{(i-1)}\right].
\end{equation}
Setting the derivative equal to zero yields a linear system in $\mathbf{x}$:
\begin{equation}
\mathbb{E}\left[\frac{\mathbf{H}^T\mathbf{(z)}\mathbf{H(z)}}{\sigma_w^2}\mid\mathbf{y};\mathbf{\hat{x}}^{(i-1)}\right]\mathbf{x} = \mathbb{E}\left[\frac{\mathbf{H(z)}}{\sigma_w^2}\mid\mathbf{y};\mathbf{\hat{x}}^{(i-1)}\right]^T\mathbf{y}.\label{eq:ml_emiteration}
\end{equation}

As is done in~\cite{WellerClassical}, the expectations in~\eqref{eq:ml_emiteration} become:
\begin{align}
\mathbb{E}\left[\frac{\mathbf{H}^T\mathbf{(z)}\mathbf{H(z)}}{\sigma_w^2}\mid\mathbf{y},\mathbf{\hat{x}}^{(i-1)}\right] &= \sum_{n=0}^{N-1}\mathbb{E}\left[\frac{\mathbf{h}_n(z_n)\mathbf{h}_n^T(z_n)}{\sigma_w^2}\mid y_n,\mathbf{\hat{x}}^{(i-1)}\right];\label{eq:ml_exphznThzn}\\
\mathbb{E}\left[\frac{\mathbf{H(z)}}{\sigma_w^2}\mid\mathbf{y},\mathbf{\hat{x}}^{(i-1)}\right]_{n,:} &= \mathbb{E}\left[\frac{\mathbf{h}_n^T(z_n)}{\sigma_w^2}\mid y_n,\mathbf{\hat{x}}^{(i-1)}\right];\label{eq:ml_exphzn}
\end{align}

The hybrid quadrature method discussed in Section~\ref{sec:background} can be used to compute the expectations in~\eqref{eq:ml_exphznThzn} and~\eqref{eq:ml_exphzn}:
\begin{align}
\mathbb{E}\left[\frac{\mathbf{h}_n(z_n)\mathbf{h}_n^T(z_n)}{\sigma_w^2}\mid y_n,\mathbf{\hat{x}}^{(i-1)}\right] &\approx \sum_{j_1=1}^{J_1}\sum_{j_2=1}^{J_2}\sum_{j_3=1}^{J_3}\frac{w_{j_1}w_{j_2}w_{j_3}\mathbf{h}_n(z_{j_3})\mathbf{h}_n^T(z_{j_3})}{{\sigma_w^2}_{j_1}p(y_n\mid\mathbf{\hat{x}}^{(i-1)})}\mathcal{N}\left(y_n;\mathbf{h}_n^T(z_{j_3})\mathbf{\hat{x}}^{(i-1)},{\sigma_w^2}_{j_1}\right);\\
\mathbb{E}\left[\frac{\mathbf{h}_n^T(z_n)}{\sigma_w^2}\mid y_n,\mathbf{\hat{x}}^{(i-1)}\right] &\approx \sum_{j_1=1}^{J_1}\sum_{j_2=1}^{J_2}\sum_{j_3=1}^{J_3}\frac{w_{j_1}w_{j_2}w_{j_3}\mathbf{h}_n^T(z_{j_3})}{{\sigma_w^2}_{j_1}p(y_n\mid\mathbf{\hat{x}}^{(i-1)})}\mathcal{N}\left(y_n; \mathbf{h}_n^T(z_{j_3})\mathbf{\hat{x}}^{(i-1)},{\sigma_w^2}_{j_1}\right).
\end{align}
Hybrid quadrature is also used to compute $p(y_n\mid\mathbf{\hat{x}}^{(i-1)})$ (see~\eqref{eq:bg_lapproxquad}). Then, the EM algorithm becomes iteratively solving~\eqref{eq:ml_emiteration} for $\mathbf{\hat{x}}^{(i)}$, using the above hybrid quadrature formulas. However, due to the three-dimensional nature of the hybrid quadrature formulas, computational cost can increase dramatically.

Due to the increased computational cost of adapting the EM algorithm to random variances, we compare the MSE performance of both EM algorithms for the same choices of parameters used in the performance plots in~\cite{WellerClassical} ($1000$ trials, $J_1 = J_2 = 9$, $J_3 = 129$). The MSE performance for both algorithms are almost identical, up to only $0.54$ dB apart. Thus, to reduce computation time when comparing performance against the Gibbs/slice sampler, the EM algorithm with known variances is used as a proxy for the EM algorithm with random variances.

\section*{Acknowledgment}

The authors thank V. Y. F. Tan for valuable discussions on Gibbs sampling and J. Kusuma for asking stimulating questions about sampling and applications of jitter mitigation. The authors also thank Z. Zvonar at Analog Devices and G. Frantz at Texas Instruments for their insights and support.

\ifCLASSOPTIONcaptionsoff
  \newpage
\fi

\ifCLASSOPTIONdraftcls
\else
	\IEEEtriggeratref{10}
\fi


\bibliographystyle{IEEEtran}
\bibliography{IEEEabrv,TSP-jittercomp-bayesian}
\end{document}